\documentclass[useAMS,usenatbib]{mn2e}

 \usepackage{graphicx}

\title[Milky Way dwarves originate from Andromeda?]{Does the dwarf galaxy system of the Milky Way originate from Andromeda?}
\author[Sylvain Fouquet et al.]{Sylvain Fouquet$^{1}$, Fran\c cois Hammer$^{2}$\thanks{E-mail: francois.hammer@obspm.fr}, Yanbin Yang$^{2,3}$, Mathieu Puech$^{2}$, Hector Flores$^{2}$\\
$^{1}$Univ Paris Diderot, Sorbonne Paris Cit\'e, GEPI, UMR 8111, F-75205 Paris, France\\
$^{2}$Laboratoire GEPI, Observatoire de Paris, CNRS-UMR8111, Univ Paris Diderot, Sorbonne Paris Cit\'e,\\ 5 place Jules Janssen, 92195 Meudon France\\
$^{3}$National Astronomical Observatoires, Chinese Academy of Sciences, 20A Datun Road, Chaoyang District, Beijing 100012, China}

\begin{document}

\date{}

\pagerange{\pageref{firstpage}--\pageref{lastpage}} \pubyear{2012}

\maketitle

\label{firstpage}

\begin{abstract}
The Local Group is often seen to be a quiescent environment without significant merger events. However an ancient major merger may have occurred in the most massive galaxy as suggested by  the M31 classical bulge and  its halo haunted by numerous stellar streams. Numerical simulations have shown that tidal tails formed during gas-rich major mergers are long-lived and could be responsible for old stellar streams and likely induce the formation of tidal dwarf galaxies (TDGs).  
Using several hydrodynamical simulations we have investigated the most prominent tidal tail formed during the first passage, which is gas-rich and contains old and metal poor stars. We discovered several striking coincidences after comparing its location and motion to those of the Milky Way (MW) and of the Magellanic Clouds (MCs). First, the tidal tail is sweeping a relatively small volume in which the MW precisely lies. Because the geometry  of the merger is somehow fixed by the anisotropic properties of the Giant Stream (GS), we evaluate the chance of the MW to be at such a {\it rendez-vous} with this gigantic tidal tail to be 5\%. Second, the velocity of the tidal tail matches the LMC proper motion, and reproduce quite well the geometrical and angular momentum properties of the MW dwarfs, i.e. the so-called disk of satellites, better called Vast Polar Structure (VPOS). Third, the simulation of the tidal tail reveals one of the formed TDG with mass and location almost comparable to those of the LMC. Our present modeling is however too limited to study the detailed interaction of gas-rich TDGs with the potential of the MW, and a complementary study is required to test whether the dwarf intrinsic properties can be accounted for by our scenario.
Nevertheless this study suggests a causal link between an expected event, an ancient, gas-rich major merger at the M31 location, and several enigma in the Local Group, namely the GS in the M31 outskirts, the VPOS almost perpendicular to the MW disk, and the presence of the MCs, two Irr galaxies near the MW.
\end{abstract}


\section{Introduction}

Despite their proximity, the properties of MW dwarf galaxies are far from being unanimously interpreted.  They are often considered as remnants of primordial galaxies that would have escaped to the numerous merger events expected in the hierarchical scenario. Within the $\Lambda$CDM paradigm they are believed to be dark matter dominated sub haloes that are gravitationally trapped in the MW halo and have not yet merged completely with their host \citep{White1978}. As the $\Lambda$CDM predicted a considerable number of massive sub-haloes relatively to the small number of observed dwarf galaxies surrounding the MW and other giant spirals, this leads to objects considerably dominated by their dark matter content. Such objects, including UFDs, are currently followed to search for dark matter annihilation, without any positive detection yet \citep[see e.g.,][]{Ackermann2012,Geringer-Sameth2011}.

However some dwarf galaxies may be tidally formed during the numerous merger events expected in the frame of the $\Lambda$CDM scenario \citep{Kroupa2012}, leading to objects mainly free of dark matter. \citet{Okazaki2000} claimed that most dwarf galaxies could be of tidal origin, although this is disputed by \cite{Bournaud2010} and by \cite{Wen2012}. A full description of the number density of TDGs is still lacking, as it has to account for gas-rich mergers that likely occurred in the past, as well as to consider a representative sampling of (major) merger orbital parameters and to describe the production and destruction rate with time of TDGs within evolving tidal tails.

A significant part of our knowledge of dwarf galaxy properties is based on the Local Group content, including irregular dwarf galaxies (dIrr) with the two Magellanic Clouds (MCs), spheroidal dwarf galaxies (dSph), and UltraFaintDwarf galaxies (UFD), which have been recently discovered \citep{Willman2005, Sakamoto2006, Zucker2006, Belokurov2006, Belokurov2007, Walsh2007}. \citet{Einasto1974} pointed out that dSphs lie close to the MW (mean distance 193 kpc), whereas the dIrr are more distant (572 kpc), except the MCs.  In order to explain this distance-dependent morphological bias, \citet{Mayer2011} suggested that the closest dwarf galaxies undergo physical phenomenons that changed their morphology from dIrr to dSph, such as stripping, stirring or tidal force.

\citet{Kalli2009} have re-estimated the proper motion of the Large Magellanic Cloud (LMC) to a significantly larger value than previous estimates, i.e. a velocity of 378 km/s relatively to the MW. Thus either MCs are passing for the first time close to the MW \citep{Besla2007} or one needs to assume a quite large dark matter content for the MW with $M_{300kpc}$= 2.7 $\pm$ 0.5 $\times$ $10^{12}$ $M_{\odot}$ \citep{Watkins2010}, i.e. a value 50 times larger than its baryonic content and even larger to that of M31. A recent measurement of the LeoII proper motion \citep{Lepine2011} provides a similar constraint, letting open the question of whether some of the MW dwarf galaxies may be unbound, and as such, would not be MW satellites.

In addition, most of the MW dwarf galaxies seems to belong to a plan-like structure, the Plane of Satellite (VPOS) which is found to be roughly perpendicular to the MW disk \citep{Kunkel1976, Lynden-Bell1976}. \citet{Kroupa2005} updated this idea by using the 11 classical dwarf galaxies and compared the satellite spatial distribution with the isotropic spatial distribution expected from $\Lambda$CDM simulations. They concluded that the dwarf galaxy distribution cannot derive from the expected $\Lambda$CDM distribution, and \citet{Metz2007a} deduced that their spatial distribution is neither spherical nor mildly prolate with a confidence of more than 99\%. Moreover, the VPOS could be a somewhat permanent structure as the angular momenta of five dSphs plus the MCs indicate a coherent motion within the VPOS \citep{Metz2008}. 

\citet{Metz2009b} summarized the different possible concepts to explain the concordance between the spatial and angular momentum distribution of the VPOS galaxies. They investigated the scenario of a dwarf galaxy group surrounding the MCs \citep{Donghia2008}, possibly coming from a filament \citep{Zentner2005,Libeskind2005}, which would be entering the MW halo in ordered motions. None of these explanations seem fully satisfactory, because it requires a quite unexpected compact group as a progenitor, and because the super galactic plane is almost perpendicular to the VPOS \citep{Metz2009b}.

A fully different alternative would be that MW dwarf galaxies originate from an ancient tidal tail \citep[see a detailed discussion on the possible tidal origin in][]{Kroupa1997,Metz2009b}, which would explain their ordered motion. In fact such an alternative was suggested quite early \citep{Lynden-Bell1976} and \citet{Pawlowski2011} have recently investigated the possibility of a major merger or a fly-by interaction (mass ratio 1:1 or 4:1) that would have occurred in the early history of the MW, producing tidal tails and then TDGs. Assuming an orbital angular momentum aligned to the VPOS, the tidal tail associated to this early interaction in the MW history would have formed TDGs that could further populate the VPOS. While this is an interesting suggestion, perhaps the MW is not the best candidate for being a merger remnant, which are generally associated with classical bulge galaxies \citep{KK2004}. Moreover, as mentioned by \citet{Pawlowski2011}, the fly-by alternative would have let the interloper well detectable, at the fringes of the Local Group\footnote{We do not consider the possibility that the interloper could be indeed the LMC, because this would require a very problematic, huge difference of cosmological growth between LMC and the MW.}.

All the above ideas are very imaginative and they may resolve the origin of the VPOS and why dwarf galaxies are part of it. However this is at the cost of introducing another ad-hoc assumption. Alternatively, we may consider the past history of the MW in the whole context of the Local Group, in which the baryonic content is dominated by M31.  Quoting \citet{VdB2005}: ``Both the high metallicity of the M31 halo, and the $r^{1/4}$ luminosity profile of the Andromeda galaxy, suggest that this object might have formed from the early merger and subsequent violent relaxation, of two (or more) relatively massive metal-rich ancestral objects". As further noticed by \citet{YY2010}, the disk plane of M31 is seen almost perpendicular, implying that tidal tails possibly formed after a gas-rich merger in the M31 history could be part of a hyper-plane that includes the MW. By tracing back in time the position of the LMC, \citet{YY2010} have shown that the LMC could be near M31, 5 to 8 Gyr ago. Moreover, \citet{Hammer2010} have proposed that M31 could be the result of a major, gas-rich merger because it provides a simple and common interpretation of most of its exceptional properties, including the Giant Stream, the outer thick disk, as well as the giant ring. Constraints from stellar population ages in these different halo substructures imply a first close passage 8 to 9 Gyr ago and a fusion time 5.5 to 6 Gyr ago \citep{Hammer2010}.

This paper intends to verify whether or not the ordered motion of the MW dwarf galaxies could be entirely due to their origin as TDGs formed from a tidal tail caused by an ancient interaction in the history of M31. The goal is to verify whether this alternative may explain two exceptional features in the Local Group: the VPOS and the MW-LMC-SMC proximity. \cite{Robotham2012} estimate the occurrence of the MW-LMC-SMC configuration for 414 MW-like galaxies in a local volume of $1.8 \times 10^{5}$ Mpc$^3$ (0.01 $\leq$ z $\leq$ 0.055). The chance to found out a galaxy having two close companions at least as massive than the SMC within a projected separation of 70 kpc and a radial separation of 400 km.s$^{-1}$ is only 0.4$\,$\% (2 examples). Moreover, in each of these configurations there is another $L^*$ luminosity galaxy within 1 Mpc from the MW-like galaxy. This may support the importance of M31 for the MCs formation, and then, accounting for the VPOS, for all the MW dwarves.

In Sect. 2, we present the VPOS properties and refine its statistical significance by including with additional constraints on the dwarf angular momenta. In Sect. 3, we describe both the physical and numerical models that we adopt for reproducing the VPOS. In Sect. 4, we try to reproduce the VPOS with a simple tidal tail toy model. Then in Sect. 5, the model is improved to closely match the work made by \citet{Hammer2010}, providing  more realistic initial conditions for the tidal tail. Sect. 6 summarizes the observational coincidences that support the M31 scenario, discusses the possible falsifications, and describes the required further studies for testing them.

In this paper, we use two coordinate systems. One is the Galactocentric coordinate which is centered on the MW \citep{VdM2002}. This system is used in Sect. 2 and Sect. 4. In Sect. 5, we use a "projection" coordinate which is centered on the current M31 position with the $z$-axis pointing from M31 to the Sun, the $x$- and $y$-axis parallel to the East and the North of celestial coordinate at M31 position. In this projection coordinate, MW is located at (x, y, z) = (6.42, -2.72, 788.85) kpc and the Sun at (0, 0, 785) kpc, assuming that the distance of M31 is 785 kpc.

\section{VPOS properties}
 
\subsection{Defining the VPOS dwarf galaxies}
\label{sec:mw_dw}

\begin{figure}
  \begin{center}
    \centering
    \includegraphics[width=1\linewidth]{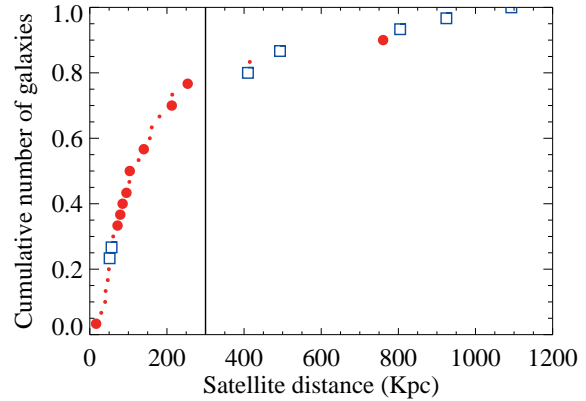}
    \caption{Distribution function of the satellite-MW distances. Large red points represent the dSphs, small red ones the UFDs, while blue squares represent the dIrrs. A line separates the MW companions considered in this study (closer than 300 kpc) from the other dwarf galaxies.}
    \label{fig:Hist_Dist}
  \end{center}
\end{figure}

The distances between the dwarf galaxies and the MW can be used as a simple criterion to verify whether dwarf galaxies are MW companions. The gravitational force of the Galaxy becomes weaker for distances larger than the virial radius, which ranges from 150 to 300 kpc. Following \citet{Metz2009a}, we consider as companions only the galaxies in a sphere with a radius of 300 kpc (see Fig. \ref{fig:Hist_Dist}). With this simple criterion, all the dIrr but the MCs are discarded, while most of the dSph and UFDs are taken into account. The next goal is to get a complete sample of dwarf galaxies within this sphere with a cut in absolute V-band magnitude. Two factors make this task difficult: the 11 classical dwarf galaxies were discovered by different sets of observations, and the Zone of Avoidance hides a part of the sky. Thus in principle, our sample may be affected by strong observational biases. The fact that the newly discovered satellites by the SDSS are systematically fainter than the classical dwarf galaxies is however reassuring.  Even Canes Vanetaci I with a $M_V$= -7.5 $\pm$ 0.5 \citep{Zucker2006} is fainter than the faintest classical dwarf, suggesting that no classical dwarf could have been missed in the sky not sampled by the SDSS \citep[80\% of the sky, see][]{Koposov2008}, except maybe in the Zone of Avoidance. Accounting for the UFDs found by the SDSS that includes a significant part of the VPOS could therefore bias the resulting statistics.

Thus, we will consider only the 11 classical dwarf galaxies grouped in a sample named ``dSphMC'': LMC, SMC, Sagittarius, Fornax, Leo I, Leo II, Sculptor, Sextans, Carina, Draco, Ursa Minor, following \citet{Kroupa2005}, and \citet{Metz2007a}. Their basic properties are listed in Table \ref{tab:data}.

\begin{table}
  \caption{Basic properties of the 11 classical dwarf galaxies. \emph{From left to right}: galactic longitude $l$, galactic latitude $b$ (both in degrees), heliocentric distance $R$ (kpc), V-band absolute magnitude $M_v$, and references: (1) = \citet{McConnachie2012}, (2) = \citet{Mateo1998}, (3) = \citet{VdB1999}}
  \begin{center}
    \begin{tabular}{lcccccr}
      \hline \hline \\
      Name     & $l$    & $b$    &     $R$      &        $M_v$     & Ref.\\ \hline
      LMC      & 280.46 & -32.88 & 50  $\pm$ 2  &  -18.1 $\pm$ 0.1 & (1), (3)\\
      SMC      & 302.80 & -44.32 & 60  $\pm$ 4  &  -16.2 $\pm$ 0.2 & (1), (3)\\
      Fornax   & 237.10 & -65.65 & 138 $\pm$ 12 &  -13.4 $\pm$ 0.3 & (1), (2)\\
      LeoI     & 225.98 & 49.11  & 250 $\pm$ 15 &  -12.0 $\pm$ 0.3 & (1), (2)\\ 
      LeoII    & 220.16 & 67.22  & 210 $\pm$ 14 &  -9.8  $\pm$ 0.3 & (1), (2)\\
      Sextans  & 243.49 & 42.27  & 86  $\pm$ 5  &  -9.3  $\pm$ 0.5 & (1), (2)\\
      Carina   & 260.11 & -22.22 & 102 $\pm$ 6  &  -9.1  $\pm$ 0.5 & (1), (2)\\
      Draco    & 86.36  & 34.72  & 82  $\pm$ 6  &  -8.8  $\pm$ 0.3 & (1), (2)\\
      UMi      & 104.95 & 44.80  & 66  $\pm$ 7  &  -8.8  $\pm$ 0.5 & (1), (2)\\
      Sgr      & 5.60   & -14.08 & 24  $\pm$ 2  &  -13.5 $\pm$ 0.3 & (1), (2)\\
      Sculptor & 287.53 & -83.15 & 87  $\pm$ 5  &  -11.1 $\pm$ 0.5 & (1), (2)\\
      \hline \\
    \end{tabular}
    \label{tab:data}
  \end{center}       
\end{table} 

\subsection{Spatial properties of the VPOS}
\label{sec:fit_DoS}

We follow the methodology of \citet{Metz2007a, Metz2009a} who have studied in detail the spatial properties of the VPOS, by fitting the position of the VPOS using a least-square method. They investigated the uncertainties associated to the direction perpendicular to the VPOS due both to the uncertainties on the distances between the MW and the satellites (using Monte-Carlo simulations) and to the fitting model (using bootstrap resampling).

\begin{figure}
  \begin{center}
    \begin{tabular}{c}
      \includegraphics[width=1\linewidth]{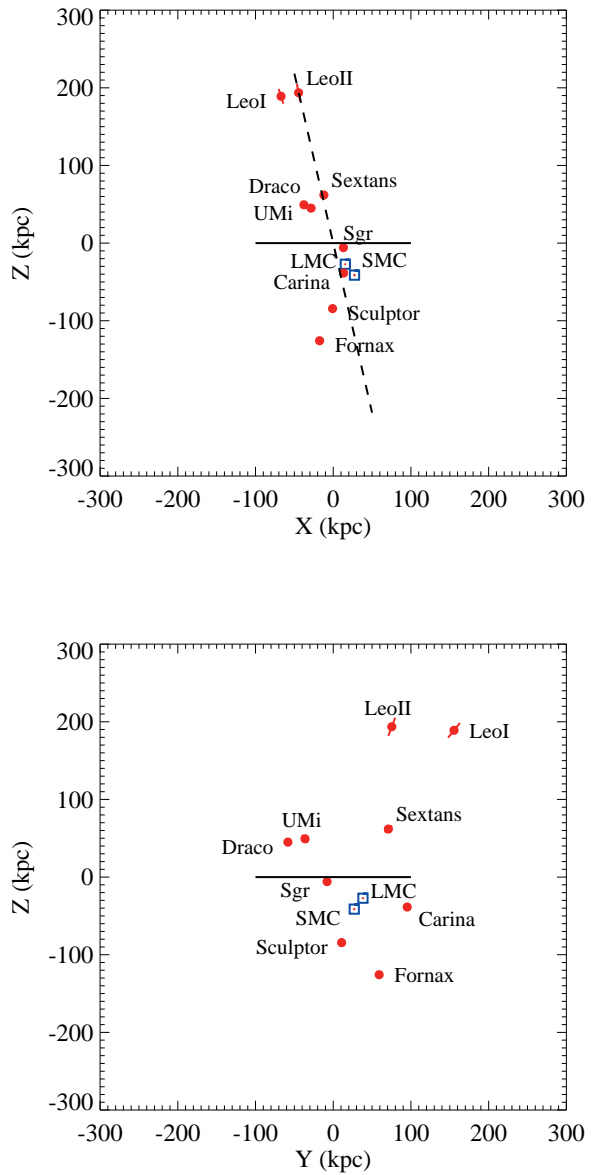}
    \end{tabular}
    \caption{VPOS position derived from the 11 classical dwarf galaxies in the MW outskirt. Red points represent the dSphs galaxies, while blue squares represent the MCs. The red lines represent the uncertainties on the heliocentric distance of the dwarf galaxies. The MW is indicated by a 100 kpc line. \emph{Top panel}: edge-on view of the VPOS (magenta dashed line). \emph{Bottom panel}: same as the top panel but rotated by 90$^{\circ}$ along the polar-axis of the MW disk.}
    \label{fig:DoS}
  \end{center}
\end{figure}

Using the data listed in Table \ref{tab:data}, we confirm that the dominant source of uncertainty is due to the fitting model (see Table \ref{tab:fitDoS1}). Figure \ref{fig:DoS} shows the resulting VPOS fit, which is found to be very similar to that derived by \citet{Metz2007a}.

\begin{table}
  \caption{Results of the VPOS fit. \emph{From left to right}:  galactic longitude and latitude of the vector perpendicular to the VPOS, $l_{VPOS}$ and $b_{VPOS}$ (deg), uncertainty due to the plane fitting method, Error1 (deg), uncertainty due to the uncertainty associated to the VPOS dwarf galaxy distance, Error 2 (deg), minimal distance from the MW center to the VPOS plan, D-center (kpc), standard deviation of the minimal distance from the VPOS dwarf galaxy position to the VPOS, thickness, T (kpc).}
  \begin{center}
    \begin{tabular}{cccccccc}
      \hline \hline \\
      $l_{VPOS}$   & $b_{VPOS}$ & Error1 & Error2 & D-Center & T    \\ \hline 
      157.4    & -12.5   & 16.4   & 1.03   & 8.07         & 18.5 \\
      \hline\\
    \end{tabular}
    \label{tab:fitDoS1}
  \end{center}
\end{table}

\subsection{Kinematic properties of the VPOS}

In addition to spatial correlations, \citet{Metz2008} have robustly confirmed that most of the individual dwarf motions lie close to the VPOS. We have extended and confirmed their study by adding two dwarf galaxies, i.e., LeoII and Sextans (see Table \ref{tab:data_vel}). In the left panel of Fig. \ref{fig:AM_obs}, we show that the angular momentum directions of all the dwarf except Ursa Minor and Sagittarius, lie close to the uncertainty range of the VPOS perpendicular direction (notice that Carina and Draco are only marginally consistent). Besides this, the Sculptor motion is found to be lying in the VPOS but counter-rotating. Indeed, its angular momentum is shifted by 180$^{\circ}$.

As the MW dwarf galaxies orbit within a thick plane that is not rotationally supported, the term Vast Polar Structure (VPOS) \citep{Pawlowski2012} is better than Disk of Satellite to describe the true nature of this structure.

\subsection{Significance of the VPOS}

We test the hypothesis of an isotropic distribution for the 10 classical dwarf galaxy positions and velocities (Leo I is discarded because its velocity is unknown). To do this, we used the statistical test proposed by \citet{Metz2007a}, which is based on fitted planes to isotropic distributions (see details in Appendix). The null hypothesis is rejected with a probability of 99.3 \% and 97.0 \% for the positions and the velocities, respectively. In an isotropic model, the position and velocity distributions are independant, which leads us to reject the isotropic distribution hypothesis in the full phase space with a probability of 99.98 \%, which further supports the significance of the VPOS.

\begin{figure*}
  \begin{center}
    \begin{tabular}{c}
      \includegraphics[width=0.5\linewidth]{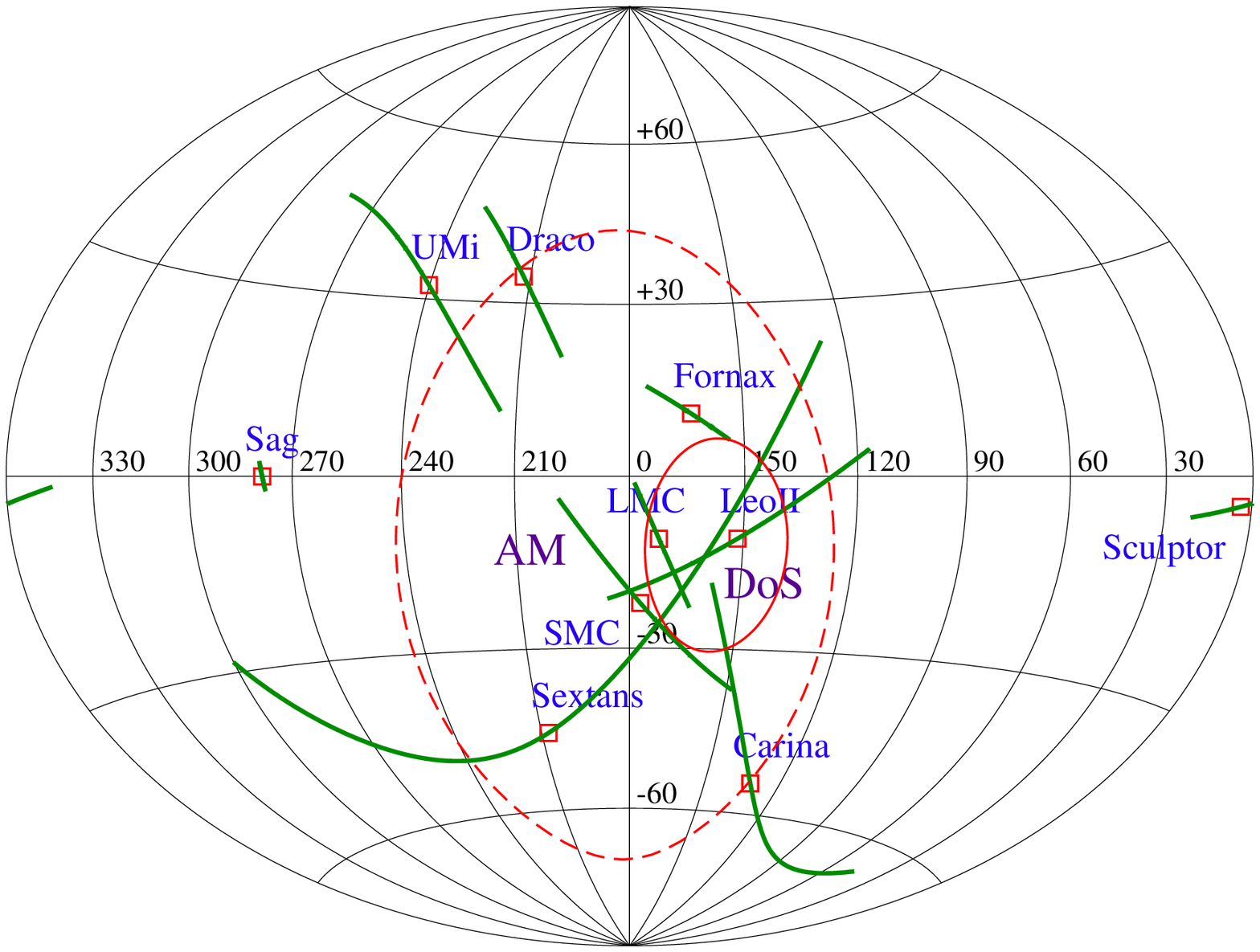}
      \includegraphics[width=0.5\linewidth]{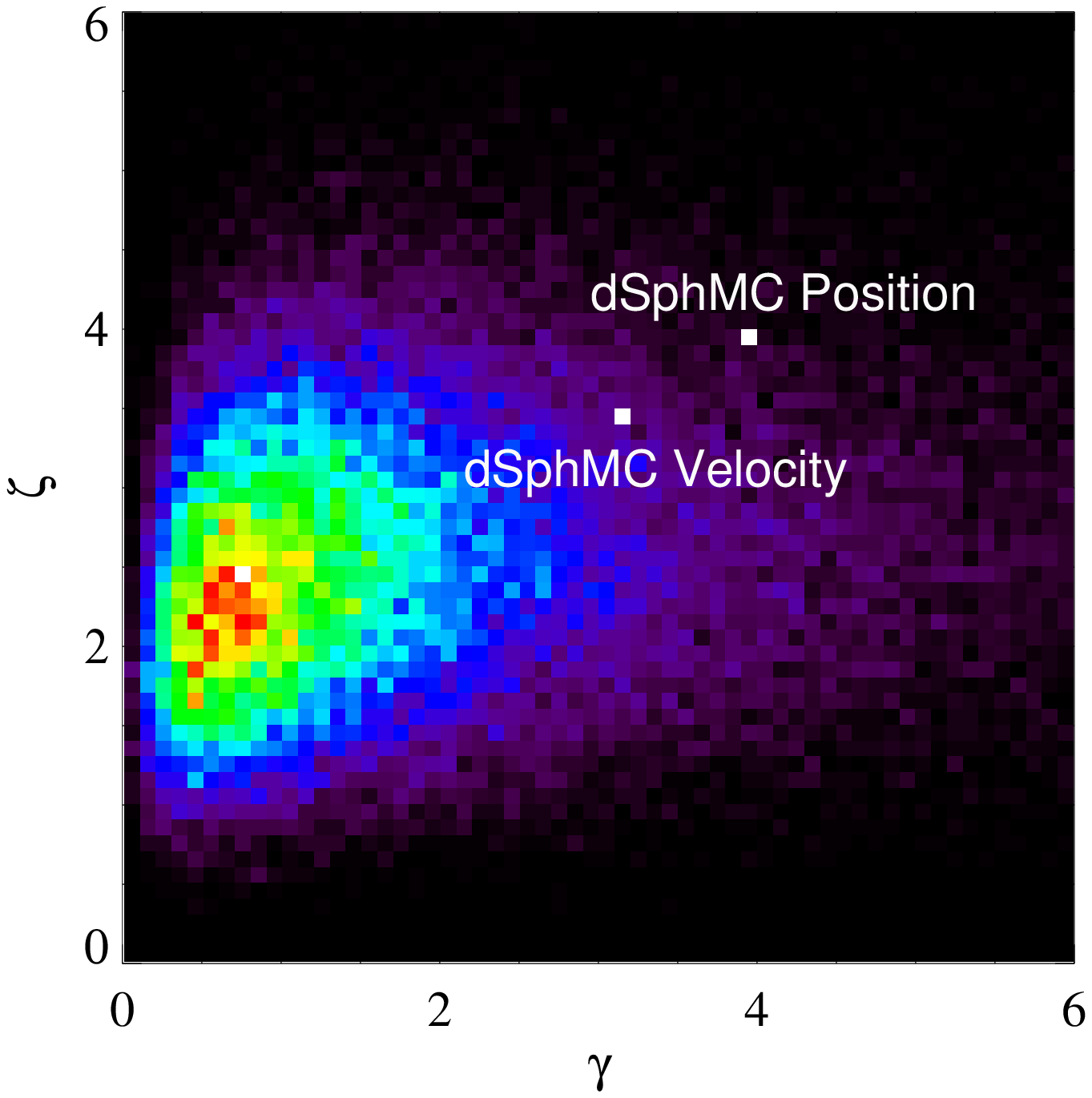}
    \end{tabular}
    \caption{
      \emph{Left panel}: Aitoff map showing the direction of the angular momenta of the classical dwarf galaxies (squares), and their uncertainties (lines). The solid ellipse shows the uncertainties associated to the direction perpendicular to the VPOS. The dashed ellipse represents the uncertainty associated to the direction of the mean angular momentum.
      \emph{Right panel}: ($\gamma$, $\zeta$) density map generated using Monte-Carlo simulations of isotropic satellite distributions (see Appendix A). A large value of $\gamma$ and $\zeta$, such as found for the MW position and velocity distributions, means that the vector perpendicular to the fitted plane have a clear preferred direction and that its uncertainty around this direction is small, which means that the fit is significant.
    }
    \label{fig:AM_obs}
  \end{center}
\end{figure*}

However more realistic cosmological simulations are providing more complex satellite geometries around their host galaxy \citep[see e.g.][]{Libeskind2005,Deason2012}. \citet{Deason2012} have used simulations that avoid an over-efficient tidal stripping of the satellites before fusion, generating a significant fraction of reconstructed spirals after major fusions \citep{Font2011}. These simulations are thus more realistic to figure out what is the distribution of satellites in host galaxies. Up to 20\% of the hosts may show a polar distribution of their satellites, while less than 10\% of the satellites may have some coherent motions. The MW satellites being both polar distributed and with coherent motions within  the VPOS, \citet{Deason2012} concluded that "if a substantial number of the classical dwarf galaxies of the Milky Way (e.g. 7 out of 10) were found to have orbital poles aligned with the normal to the disc of satellites, then this would be inconsistent with the results of our simulations". As shown by \citet{Metz2007a} this is because the VPOS is particularly thin which requires the accretion of a large, hypothetic group of dwarf galaxies that is otherwise naturally brought by remnant tidal tails. Thus additional data on the proper motions are still required, though the present data favor an exceptional MW satellites alignment of their locations and motions (see Fig. \ref{fig:DoS}, \ref{fig:AM_obs}).

\begin{table*}
  \caption{Proper motions of 10 classical dwarf galaxies. \emph{From left to right}: Radial velocity (km/s), proper motion (mas/yr), both in the heliocentric restframe, total velocity, radial velocity, transverse velocity (all in km/s in the galacto-centric restframe), and references: (1) = \citet{Vieira2010}, (2) = \citet{Piatek2007}, (3) = \citet{Lepine2011}, (4) = \citet{Walker2008}, (5) = \citet{Piatek2003}, (6) = \citet{Scholz1994}, (7) = \citet{Piatek2005}, (8) = \citet{Ibata1997}, (9) = \citet{Piatek2006}.}
  \begin{center}
    \begin{tabular}{lccccccr}
      \hline \hline \\
      Name     & $V_{rad}$ &  $\mu_{\delta}$  & $\mu_{\alpha}\cos(\delta)$ & $V_{tot}$ & $V_{rad}$  & $V_{trans}$ & Ref. \\ \hline
      LMC      &   278     &  0.39 $\pm$0.27  & -1.89 $\pm$0.27            & 347$^{ +73}_{ -69}$ &  103$^{+11}_{-11}$  &  331$^{ +73}_{ -69}$ & (1) \\
      SMC      &   158     & -1.01 $\pm$0.29  &  0.98 $\pm$0.30            & 221$^{+110}_{-120}$ &   29$^{+14}_{-14}$  &  219$^{+110}_{-121}$ & (1) \\
      Fornax   &    53     & -0.36 $\pm$0.041 &  0.476$\pm$0.046           & 188$^{ +38}_{ -39}$ &  -31$^{ +2}_{ -2}$  &  185$^{ +38}_{ -39}$ & (2) \\
      LeoII    &    79     & -0.033$\pm$0.151 &  0.104$\pm$0.113           & 268$^{+171}_{-184}$ &   22$^{ +5}_{ -5}$  &  267$^{+172}_{-186}$ & (3) \\
      Sextans  &   224     &  0.10 $\pm$0.44  & -0.26 $\pm$0.41            & 260$^{+211}_{-182}$ &   88$^{+20}_{-20}$  &  241$^{+216}_{-230}$ & (4) \\
      Carina   &   229     &  0.15 $\pm$0.09  &  0.22 $\pm$0.09            & 131$^{ +64}_{ -60}$ &   27$^{ +4}_{ -4}$  &  128$^{ +65}_{ -61}$ & (5) \\
      Draco    &  -292     &  1.1  $\pm$0.3   &  0.6  $\pm$0.4             & 563$^{+174}_{-180}$ &  -62$^{+12}_{-12}$  &  559$^{+176}_{-184}$ & (6) \\
      UMi      &  -247     &  0.22 $\pm$0.16  & -0.50 $\pm$0.17            & 161$^{ +52}_{ -44}$ &  -76$^{ +8}_{ -8}$  &  141$^{ +61}_{ -58}$ & (7) \\
      Sgr      &   140     & -0.88 $\pm$0.08  & -2.65 $\pm$0.08            & 303$^{ +10}_{ -10}$ &  138$^{ +1}_{ -1}$  &  269$^{ +11}_{ -11}$ & (8) \\
      Sculptor &   110     &  0.02 $\pm$0.13  &  0.09 $\pm$0.13            & 227$^{ +65}_{ -65}$ &   79$^{ +6}_{ -6}$  &  212$^{ +69}_{ -72}$ & (9) \\
      \hline \\
    \end{tabular}
    \label{tab:data_vel}
  \end{center}
\end{table*}

\section{Forming the VPOS as a remnant of a tidal tail}

\subsection{Physical model}
\label{sec:model}

\citet{Hammer2010} have proposed that a gas-rich major merger occurred during the history of M31. They used an N-body, hydrodynamical simulation \citep[GADGET2,][]{Springel2005} to reproduce most of the M31 structures such as the bulge, the thin and thick disks, the Giant Stream, and the 10 kpc ring. In their model, the first passage between the galaxies occurred more than 8 Gyr ago, and the fusion time 2.5-3 Gyr later. A significant part of the matter ejected during the first-passage lies in a tidal tail. As we will show in Sect. \ref{subsec:TT_model}, for most of the models reproducing M31, this tidal tail is found to pass close to the MW.

Several observations and simulations have suggested that dwarf galaxies, named Tidal Dwarf Galaxies (TDGs), can form in such tidal tails \citep[see][and references therein]{Okazaki2000}. \citet{Bournaud2006} have run 96 major merger simulations in order to better understand TDG properties and their formation mechanisms. Most of their simulations revealed the formation of TDGs with stellar masses larger than 10$^8$ M$_{\odot}$. However, amongst the MW companions, 9 of the 11 classical dwarf galaxies have stellar masses ranging from $\sim$ 10$^5$ to $\sim$ 10$^7$ M$_{\odot}$ \citep{Strigari2008}. Using simulations with higher resolution, \citet{Bournaud2008} have further shown that young stellar objects, star clusters, and TDGs with masses ranging from 10$^5$ to 10$^9$~M$_{\odot}$ can also be reproduced. All these simulations suggest that it is possible to form TDGs with a relatively large range of mass, as a result of a major merger event.

In the following sections, we investigate whether the remnant of such a tidal tail could be at the origin of the VPOS dwarf galaxies. \citet{Pawlowski2011} were the first to study in detail such a hypothesis. They simulated the evolution of tidal tails created by major mergers or fly-byes in the MW past, $\sim$ 10 Gyr ago. They explored the orbital parameters of the progenitors to match the properties of the VPOS. However, a fly-by seems unlikely, since no other massive galaxy but M31 is observed in the Local Group. In the present study, we investigate the formation of the VPOS as a result of a tidal tail ejected during a merger that occurred in the past history of M31 rather than in that of the MW. Indeed, \citet{YY2010} have shown that the position of the LMC could be traced back toward M31, $\sim 6$ Gyr ago, assuming an appropriate choice for the M31 tangential velocity and the MW halo shape. This suggests that all VPOS dwarf galaxies could have formed as TDGs resulting of a major merger event in the M31 history.

\subsection{Numerical models}
\label{sec:num_model}

To investigate this scenario, we developed a C code that simulates the trajectories of tidal tails populated with TDGs. They are formed and ejected during a major merger, and eventually interact with the MW.

The small mass of the TDGs, as well as the large distance between M31 and the MW, make the tidal effects undergone by the two spiral galaxies negligible. This allows us to assume that their gravitational potentials are static, and to model them analytically. Following \citet{YY2010}, we assume Navarro-Frenk-White halos \citep{NFW1997}, Hernquist bulges \citep{Hernquist1990}, and exponential disks \citep{YY2010} for M31 and the MW, while dwarf galaxies and the LMC are assumed to be point masses.

In the simulations, we first trace back the position of the LMC, the MW, and M31 backwards in time until the LMC reaches the outskirt of M31. The unknown tangential velocity of M31 is taken as a free parameter, while other parameters are listed in Table \ref{tab:data} \& Table \ref{tab:data_vel}. Since the LMC is assumed to be a TDG remnant, its mass is just equal to its baryonic mass, $\sim 3.10^9$ M$_{\odot}$. We have verified that the dynamical friction is negligible in the computation of the LMC trajectory.

Second, we used numerical models of tidal tail formation to set up initial conditions for the positions and velocities of the TDGs in the surrounding of M31. The position of the LMC back in time in the surrounding of M31 is used to constrain the tidal tail properties. Finally, we let evolve the TDGs until present-time. At the end of the simulation, we analyze the TDG positions and angular momenta, and test whether they match the present-day VPOS properties.

\section{A tidal tail toy model constrained by the LMC trajectory}
\label{sec:mod1}

\subsection{Initial conditions for the TDGs}

The goal of the proposed toy model is to verify whether or not the gross properties of the VPOS can be described by particles linked to the over-densities found in tidal tails. As described in Sect. \ref{sec:num_model}, we first trace back in time the position of the LMC, MW, and M31, 5.5 Gyr ago. This corresponds to the epoch when the LMC was found to be at 50 kpc from M31 according to \citet{YY2010} (see their Table 3). To be consistent with their study, the MW and M31 total masses are respectively choosen to be 10$^{12}$ M$_{\odot}$ and $1.6 \times 10^{12}$ M$_{\odot}$, while the LMC velocity is taken from the study made by \citet{Kalli2009}, i.e., V = 384 km/s.

\begin{figure}
  \centering
  \includegraphics[width=0.65\linewidth]{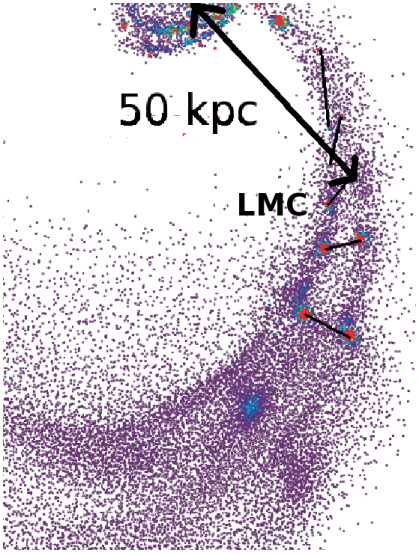}
  \caption{Trajectories of five over-densities (black lines) in a tidal tail during a time step of 262 Myr \citep{Wetzstein2007}. The third over-density is assumed to represent the position of the LMC back in time. The distance between the LMC and the merger remnant center is 50 kpc.}
  \label{fig:TT_model}
\end{figure}

The position and velocity profiles of the tidal tail are extracted from the work of \citet{Wetzstein2007}. Indeed, they studied the formation of TDGs produced in a major merger and estimated the effects of initial conditions on the TDG properties. Two snapshots separated in time by 262 Myr are extracted from their simulation EG2 (see Fig. \ref{fig:TT_model}). Five density peaks are found to be prominent within the tidal tail. We assume that the third peak represents the LMC location in the past. This arbitrary choice implies that the part of the tidal tail which is located farther than the LMC position will escape from M31, which is quite expected during the typical evolution of a tidal tail.

We use the relative positions of the five prominent peaks to set up initial positions of five groups of TDGs. Each group, even that with the LMC, is assumed to contain six TDGs, which are modeled as point masses with a total mass equals to 10$^8$ M$_{\odot}$, i.e., close to that of the SMC. We choose six particles per peak in order to populate enough the outskirts of the MW at the end of the simulation. The mass distribution in each group of particles is assumed to follow a Plummer density mass profile with a typical radius of 10 kpc, which constrains the internal velocity dispersion to be $\sim$ 16 km/s in order to have a static group. The bulk velocity of the TDG groups is roughly estimated by comparing the positions of the peaks in the two snapshots (see Fig. \ref{fig:TT_model}). The tidal tail plane that encompasses the positions and motions of the TDGs is rescaled, rotated, and translated in order for the third density peak to match both the position and velocity of the LMC, 5.5 Gyr ago.

\subsection{The results}
\label{sec:mod1_res}

A set of 100 simulations has been run by randomly generating internal position and velocity for the TDGs in the five groups, consistent with a Plummer model. For each simulation, we only kept particles for which the trajectories end within 300 kpc from the MW centre, i.e., the distance limit used in Sect. \ref{sec:mw_dw} to define a dwarf galaxy as a MW companion. The VPOS properties and the mean angular momentum direction are calculated following the same method that the one used for the observations (see Sect. \ref{sec:fit_DoS}). The resulting plane thickness distribution for the 100 simulations has an average of 18.3$\pm \, 6$ kpc, which is consistent with the observed value (see Table \ref{tab:fitDoS1}).

\begin{figure}
  \centering
  \includegraphics[width=0.9\linewidth]{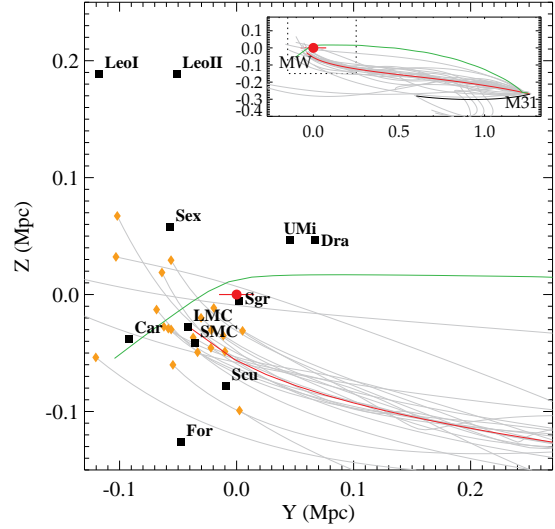}
  \includegraphics[width=0.9\linewidth]{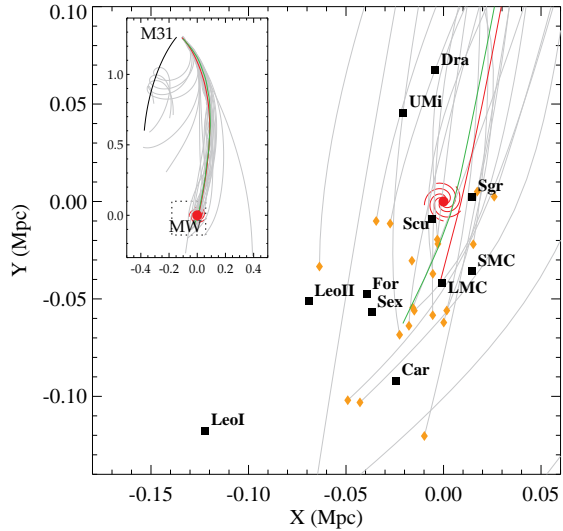}
  \caption{Comparison between the positions of the simulated TDGs (diamonds) and the observed satellites (squares). The upper and lower panels show the projection in Y-Z and X-Y galactic coordinate plane, respectively. The MW is represented as a red spiral. The grey lines represent the trajectories of each simulated TDG. The in-set panels show a global view of the trajectories of the simulated TDGs (grey lines) as well as that of M31 (black line). The trajectory of the LMC is also shown as a red line. The green line illustrates a counter-rotating trajectory, which may explain the motion of Sculptor.}
  \label{fig:YY_pos}
\end{figure}

Figure \ref{fig:YY_pos} shows typical trajectories of simulated TDGs. Particles are found to travel on both sides of the MW disk plane, although most of them, including the LMC, pass below it. There are some particles passing above the MW disk plane, with an angular momentum direction shifted by $180^{\circ}$ relative to the simulated VPOS direction. This configuration is similar to the counter-rotation of Sculptor in the MW environment.

\begin{figure*}
  \centering
  \includegraphics[width=0.9\linewidth]{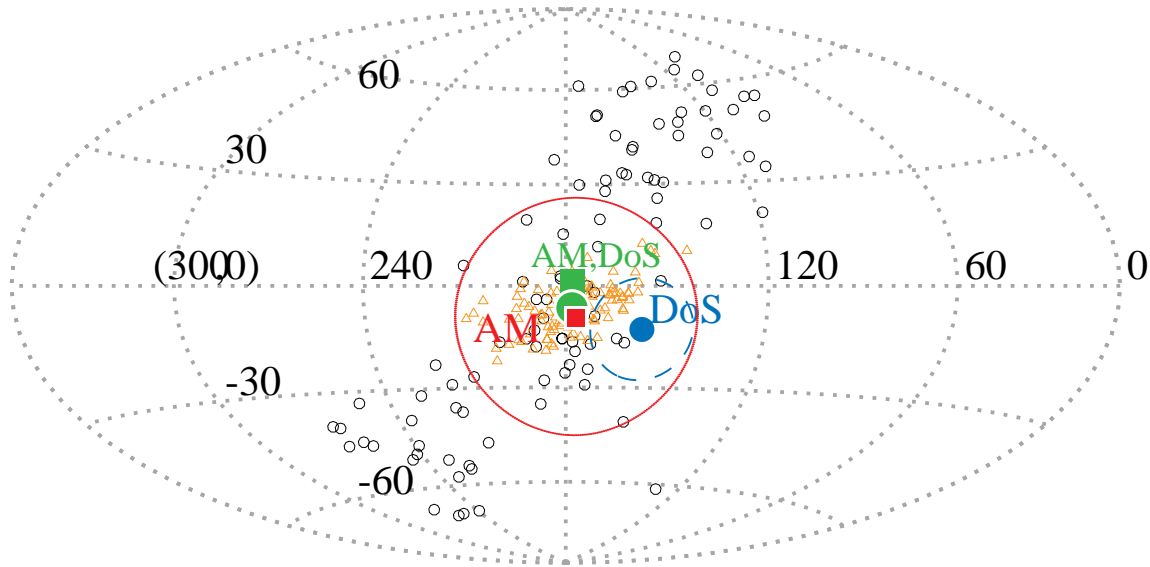}
  \caption{Aitoff map comparing the VPOS and angular momentum (AM) directions between simulations and observations. The red square and the blue dots indicate the direction of the mean AM and the direction of the VPOS, respectively, as derived from observations \citep{Metz2007a,Metz2008}. Their uncertainties are represented by the red circle (AM) and by the blue dashed circle (VPOS). The open triangles and circles represent the directions of the VPOS and the mean AM, as resulting from 100 simulations (see Sect. \ref{sec:mod1_res}). Their mean values are indicated by a green square (VPOS) and a green dot (AM).}
  \label{fig:YY_AM}
\end{figure*}

In Fig. \ref{fig:YY_AM}, we plot for the 100 simulations the resulting mean angular momentum directions and the directions perpendicular to the simulated VPOS. These directions populate the region occupied by the uncertainty range corresponding to the observed dwarf galaxies. We find that the mean angular momentum direction of the 100 simulations is consistent within uncertainties with the mean direction of the observed angular momentum. The mean perpendicular direction for the 100 simulated VPOS is also close to that derived from observations, although not strictly within the uncertainty range. However, for some simulations, the perpendicular direction is encouragingly found to lie within this range.

This very simple model illustrates how TDGs associated with a tidal tail can easily reproduce most of the VPOS properties. By varying the radius of the Plummer model and the LMC position in the tidal tail, one can reproduce the fraction of dwarf galaxies having the same angular momentum direction that the LMC (8 dwarf galaxies) and those having it inverted as Sculptor, i.e. a very similar optimization that done by \citet{Pawlowski2011}.

It is clear that such a model is quite ad-hoc since the dominant motion of the simulated TDGs is that of the LMC, which is a part of the VPOS. The LMC velocity measurement used is larger than that found by \citet{Vieira2010}. Using their value would increase the trace back time needed by the LMC to reach M31 by more than 2.5 Gyr. This time could match the formation epoch of the first tidal tail according to the model of \citet{Hammer2010} described in Sect. \ref{sec:model}, i.e 8.5 Gyr ago. Moreover, the TDGs of the VPOS remain too close to the LMC and thus do not populate enough the region above the MW in Fig. \ref{fig:YY_pos}. This shows the limit of this ``toy model''. Finally, such a toy model, as well as that proposed by \citet{Pawlowski2011}, do not allow us to distinguish the origin of the tidal tail (i.e, the MW or M31). These remarks lead us to make a more realistic model of the tidal tail. In the next section, we propose to link far more robustly the modeling of M31 and its halo with simulations testing the VPOS formation. In principle it should bring a number of constraints large enough to rule out or support the M31 tidal tail hypothesis.

\section{A tidal tail generated by an ancient merger at the M31 location}
\label{sec:mod2}

\subsection{M31 modelling as a remnant of a major merger}
\label{subsec:TT_model}

\citet{Hammer2010} have run a series of gas-rich, major merger simulations designed to reproduce both the M31 galactic substructures and the extended structures in its outskirts. The orbital geometry of the interaction is close to be polar, driven by the need to reproduce the giant 10 kpc ring, while  stellar ages in the M31 halo substructures constrain the pericenter radius. In this family of models, two tidal tails systematically form during the first passage, as well as additional ones during the second passage and/or fusion. Each tidal tail is made by material stripped from each M31 progenitor, resulting in an angular momentum that combines that of the corresponding galaxies with the orbital angular momentum, which is the dominant component. We have tested a significant series of models from  \citet{Hammer2010} (see their Table 3) and found that the first-passage tidal tail, associated to the minor encounter, passes systematically close to the MW, i.e., generally at 50 kpc and always at less than 200 kpc (see Fig. \ref{fig:TT_simu}). In other words, the MW lies within the past M31 merger orbital plane, and the large area swept by the tidal tail motion, leads to a significant chance for a tidal tail-MW interaction. In fact the first-passage tidal tail is the most extended one. It contains the largest amount of baryons expelled during the merger, and the largest amount of gas, and the lowest metal abundance compared to other tidal tails. In the \citet{Hammer2010} model, the Giant Stream is linked to the tidal tail formed during the second passage, and caused by stars returning to the remnant galaxy leading to the formation of loops \citep[see their Fig. 8 as well as a complete description of the loop mechanism in][]{Wang2012}.

The model used in the present study is a refined version of  \citet{Hammer2010}  (10th column in their Table 3) with a slightly higher mass ratio (3.5 instead of 3), and an improvement of the initial conditions, i.e., the two interlopers at $t = 0$ are separated by 200 kpc instead of 80 kpc. The most significant change is due to the use of a larger number of particles (2.4 millions instead of  540 thousands), which allows us to generate gravitational instabilities within the tidal tail, under the assumption of a softening length of 0.1 and 0.12 kpc for baryons and dark matter, respectively. A more subtle change is about the Giant Stream modeling, which is associated to the second order loop instead of the first order loop \citep[see Fig. 8 in][]{Hammer2010}. This results in a rotation within the orbital plane by approximately 50 degrees, which only slightly impacts the relative angle between the first tidal tail and the assumed Giant Stream. This updated model now reproduces both the Giant Stream and the Northern loop discovered by the PAndAS team \citet{Mackey2010}, those being related to material stripped in the loop formed at the fusion and at the first passage, respectively (see details in Yang et al., 2012, in prep.).

\subsection{Initial conditions for the TDGs}

Instead of relying on the simulations of \citet{Wetzstein2007} as in Sect. \ref{sec:mod1}, we now use the \citet{Hammer2010} merger model for M31 (as updated above) to setup initial conditions for the position and velocity of the TDGs in the M31 outskirts.

We adopt a baryonic-to-total mass ratio of 20\% as used by \citet{Hammer2010} instead of 6\% as in  \citet{YY2010}, as well as in Sect. \ref{sec:mod1}. Following \citet{YY2010}, the dark matter scale length is tuned to fit the rotation curve. A total mass of $5.5 \times 10^{11}$ M$_{\odot}$ ($3.3 \times 10^{11}$ M$_{\odot}$) provides a $R_{vir} = 70$ kpc and $c = 10$ ($R_{vir} = 60$ kpc and $c = 6$) for M31 (the MW). This affects the \citet{YY2010} value for the M31 tangential velocity that is required for the LMC to be near M31 in the past, leading to ($v_{X}$, $v_{Y}$) equals to (-135, 10) km/s ($\mu_W = -71\, \mu$as.yr$^{-1}$, $\mu_N = -21\, \mu$as.yr$^{-1}$) instead of (-105, -7) km/s ($\mu_W = -62\, \mu$as.yr$^{-1}$, $\mu_N = -25\, \mu$as.yr$^{-1}$). \citet{YY2010} show that such values for the M31 tangential velocity are comparable to the radial velocity of M31 (129 km/s) and explain why they can be substantially larger than the tentative estimates made by \citet{VdM2008}. Our values (and Yang \& Hammer values) are consistent within 1.7 (1.1) $\sigma$ from the recent evaluation of \cite{VdM2012} using very small proper motions of few M31 stellar fields based on a 5 year time basis.

We first verify whether the adopted simple spherical potential for M31 in the analytic model is consistent with \citet{Hammer2010} simulations. After examining the dark matter distribution in the \citet{Hammer2010} simulations, we choose to pick up the tidal tail one Gyr after fusion, when the mass is relatively stable and spherically distributed. At this epoch, the MW is too far from M31 to affect the tidal tail. As this model does not allow us to follow the internal evolution of the tidal tail, we simplify the code by neglecting the force between the tidal tail particles. For each tidal tail particle the simulation is therefore a 3-body problem. In order to check whether the tidal tail has the same evolution in the GADGET2 code used by \citet{Hammer2010} that in the 3-body code used here, we ran two simulations with the same initial conditions. After 5 Gyr, the tidal tail positions and velocities are similar for the two simulations within 5\%, which confirms that our choice for the initial conditions does not impact the results significantly.

The different stages of the simulation are those described in Sect. \ref{sec:num_model}. The tidal tail is selected from the GADGET simulation, one Gyr after fusion time, and inserted into the 3-body problem. This corresponds to 4 Gyr after the first passage. As the first passage is estimated to occur 8.5$\pm 0.5$ Gyr ago \citep{Hammer2010}, this corresponds to a look back time of 4.5$\pm 0.5$ Gyr.

\begin{figure}
  \centering
  \includegraphics[width=1\linewidth]{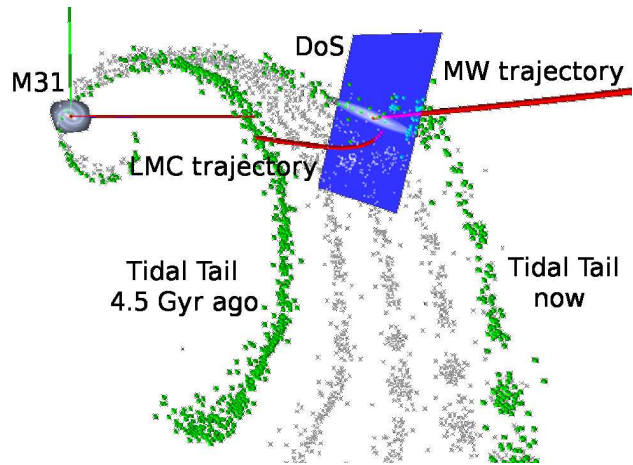}
  \caption{Schematic illustration of the VPOS formation as a result of a tidal tail ejected by the M31 major merger. The tidal tail sweeps space for 4.5 Gyr followed by the LMC and finally passes through the MW trajectory. The green dots represent the initial and final position of the tidal tail, while white dots represents intermediate steps. The M31 and MW disks are represented with a size of 100 kpc, as well as the observed VPOS (blue rectangular) and the y- and z-axis of the M31 rest-frame (green and red lines).}
  \label{fig:TT_IC}
\end{figure}

\begin{figure*}
  \begin{center}
    \begin{tabular}{ccc}
      \includegraphics[width=0.2\linewidth]{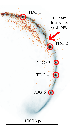} &
      \includegraphics[width=0.35\linewidth]{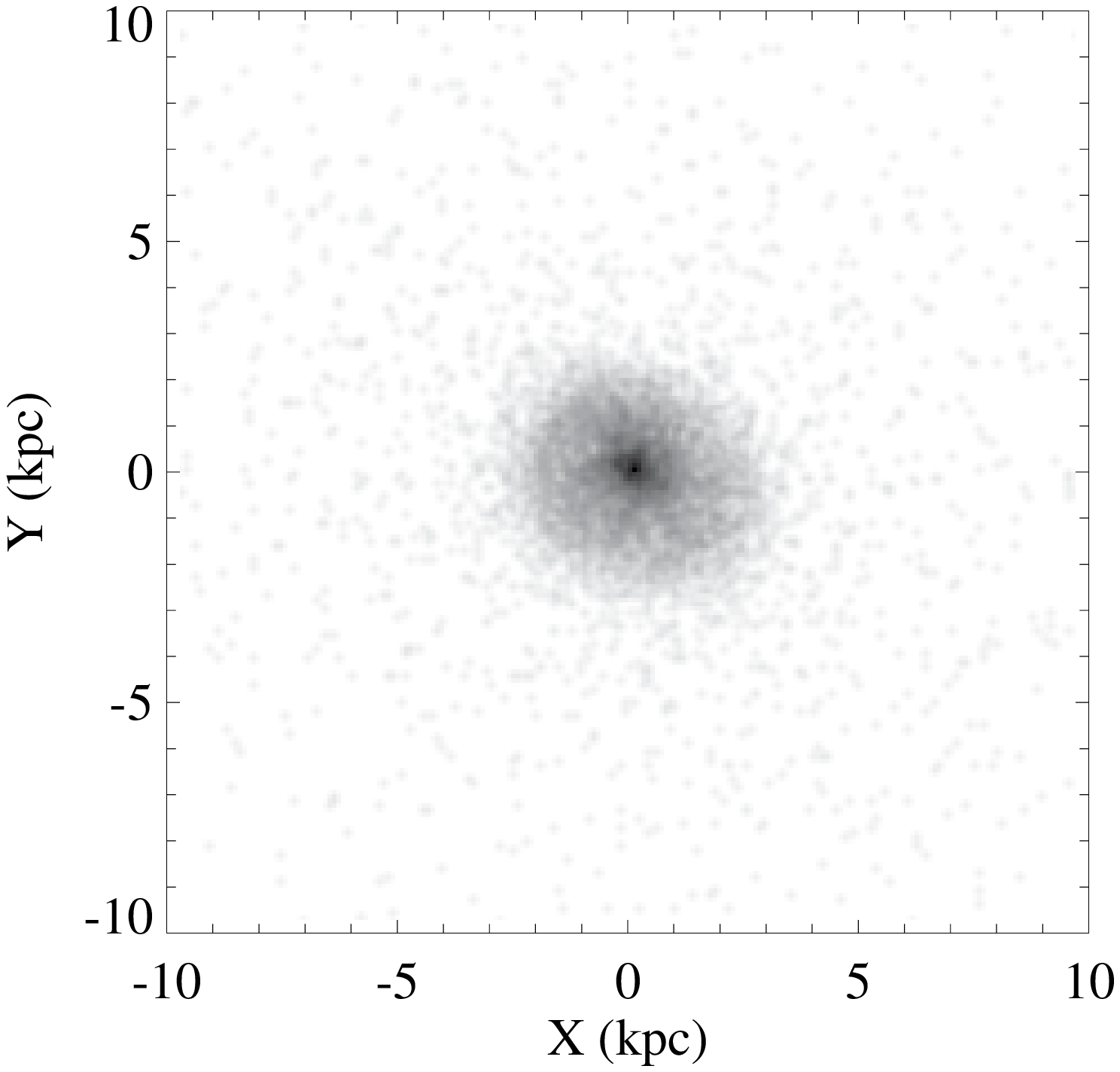} &
      \includegraphics[width=0.45\linewidth]{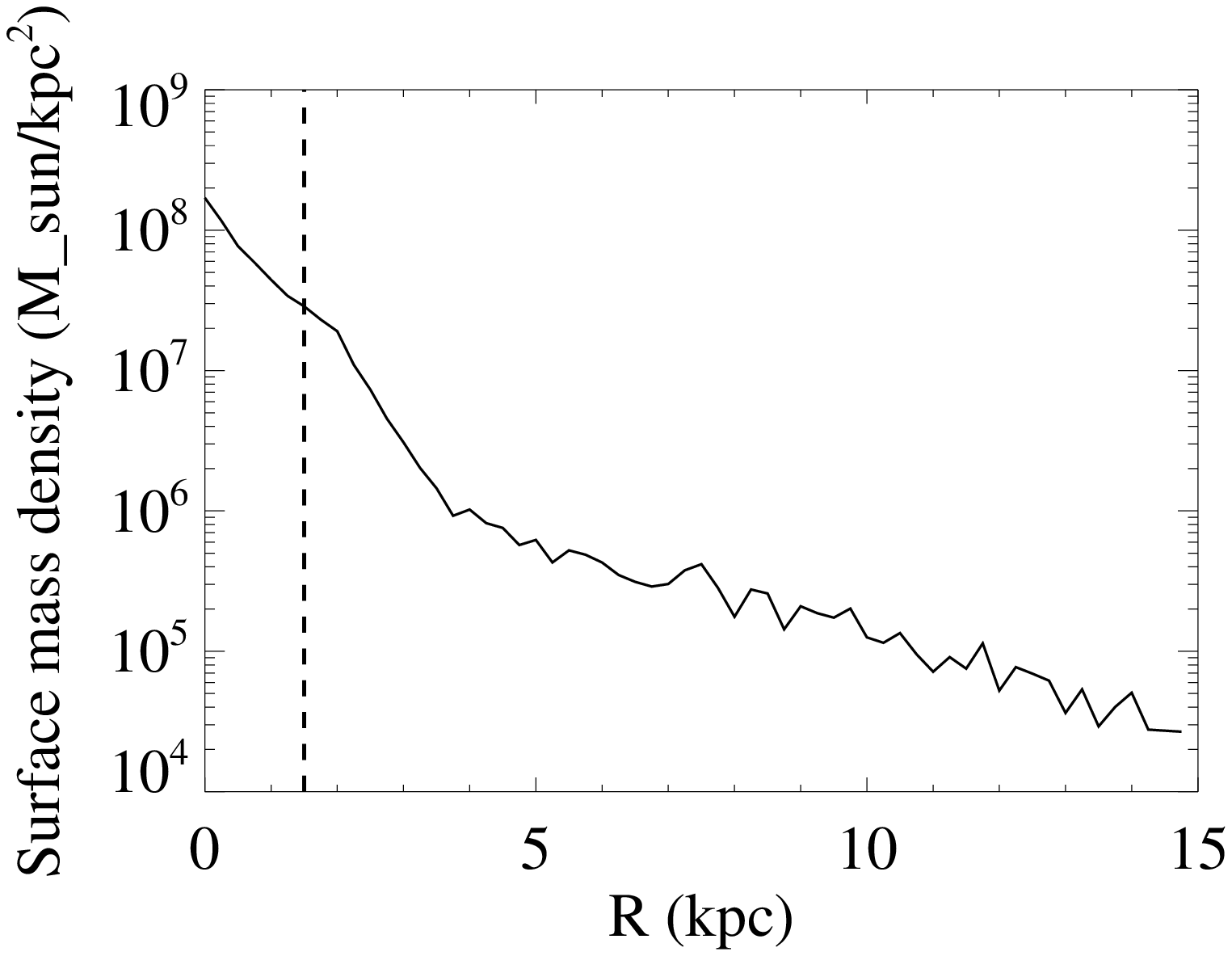} \\
    \end{tabular}
    \caption{\emph{Left panel}: Gas density map of the first tidal tail, 9 Gyr after the first passage of the merger. Gas density is coded using blue-grey color, while stellar density is shown in orange. Five red circles indicate the location of the five most massive TDGs (mass larger than $\sim$ 10$^8$ M$_{\odot}$), which are labeled TDG1 to TDG5. \emph{Middle panel}: Close-up of the TDG2. \emph{Right panel}: Radial profile of the mass surface density of TDG2. The half mass radius $\sim$ 1.5 kpc is indicated as a dashed line.}
    \label{fig:TT_simu}
  \end{center}
\end{figure*}

\subsection{Simulated VPOS properties}
\label{sec:results}

The conditions for reproducing the VPOS using a merger model for M31 are considerably more constrained than in Sect. \ref{sec:mod1} because of the required match between the tidal tail velocity and that of the LMC. In fact, we realize that for most models in  \citet{Hammer2010} such a condition is verified as illustrated in Fig. \ref{fig:TT_IC}. Besides the choice of the model reproducing the most accurately the properties of M31 (Yang et al., 2012, in prep.), the location of the tidal tail can be fine-tuned using a rotation by less than $\pm$ 20 degrees along the M31 rotational axis \citep[see][]{Hammer2010}. Such a fine-tuning preserves the modeling of the Giant Stream and is required for matching the look-back time at which both the LMC and tidal tail are found at a common position in the M31 outskirts. Figure \ref{fig:TT_IC} shows the evolution of the tidal tail together with the LMC, and the MW trajectories relative to M31. It illustrates that for a given set of transversal velocities for the MW relative to M31, the LMC is found to travel between the two galaxies. Without more than fine tuning, to have its trajectory matched by the predicted tidal tail formed after the first encounter between the two progenitors of M31.

\begin{figure*}
  \begin{center}
    \begin{tabular}{c}
      \includegraphics[width=1.0\linewidth]{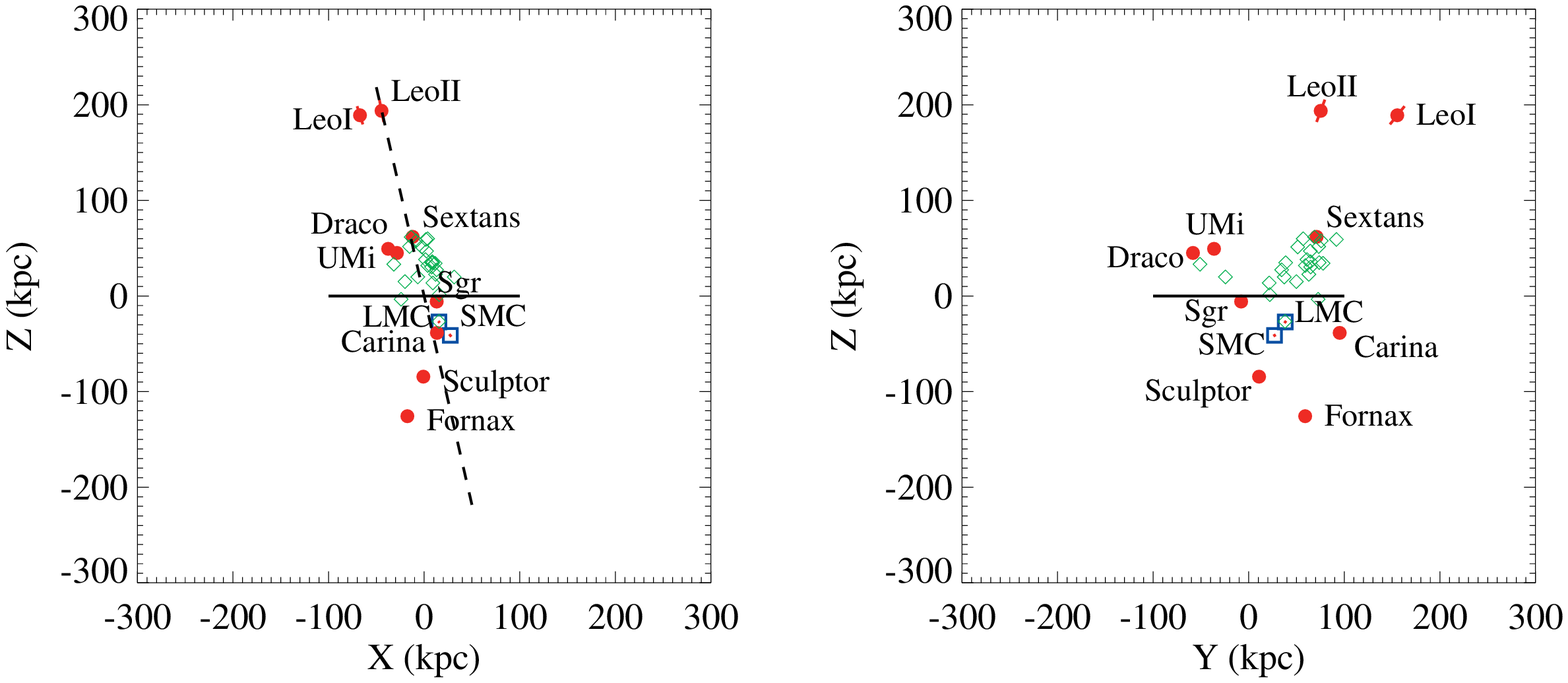} \\
      \includegraphics[width=1.0\linewidth]{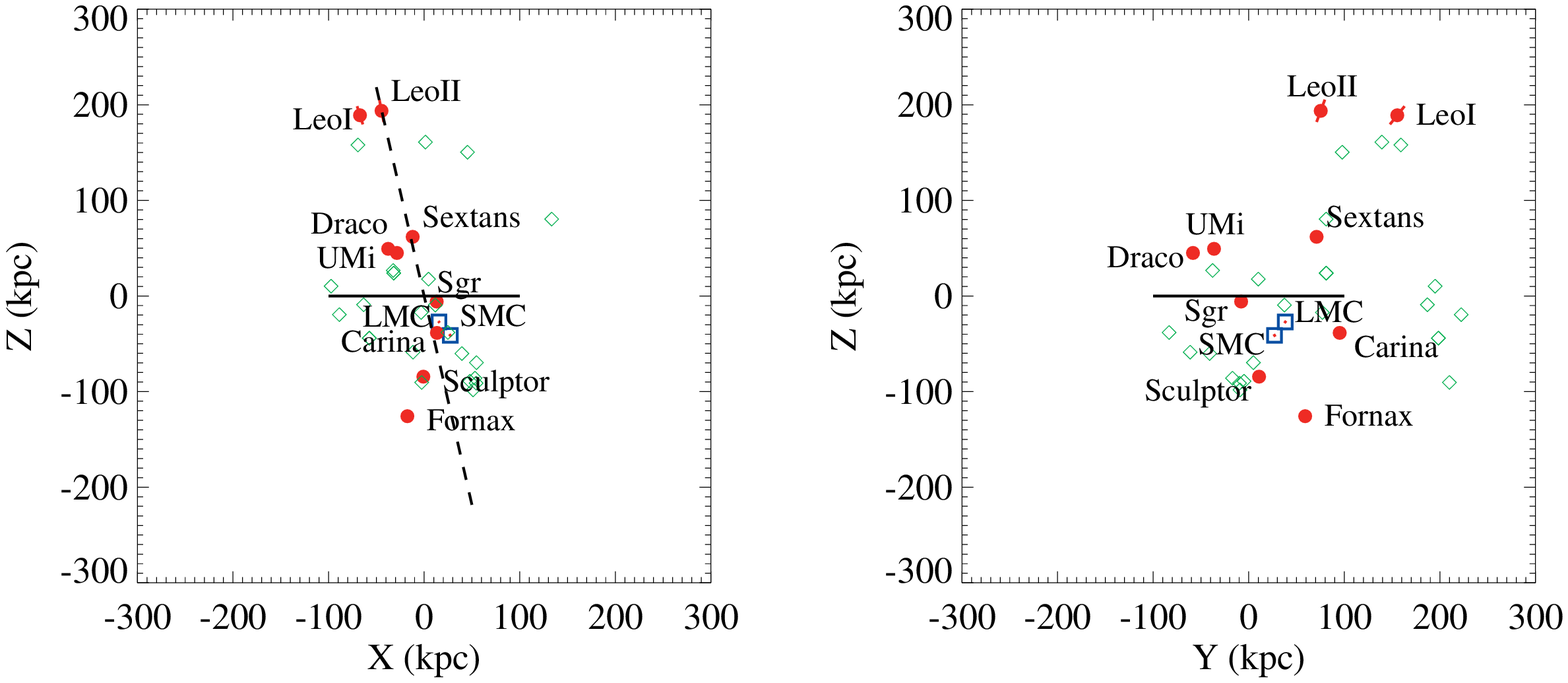} \\
    \end{tabular}
    \caption{ Face-on (\emph{left panels}) and edge-on (\emph{right panels}) views of the observed VPOS populated by the tidal tail particles (green diamonds). The 11 classical dwarf galaxies are represented by red points (dSph) and blue open squares (MCs). The \emph{top panel} corresponds to the tidal tail with a small extension and the \emph{bottom one} to the extended tidal tail. For both, there are 25 particles randomly selected into the clumps which interact with the MW (green diamonds).
    }
    \label{fig:DoS_simu}
  \end{center}
\end{figure*}

\begin{figure*}
  \begin{center}
    \begin{tabular}{c}
      \includegraphics[width=0.7\linewidth]{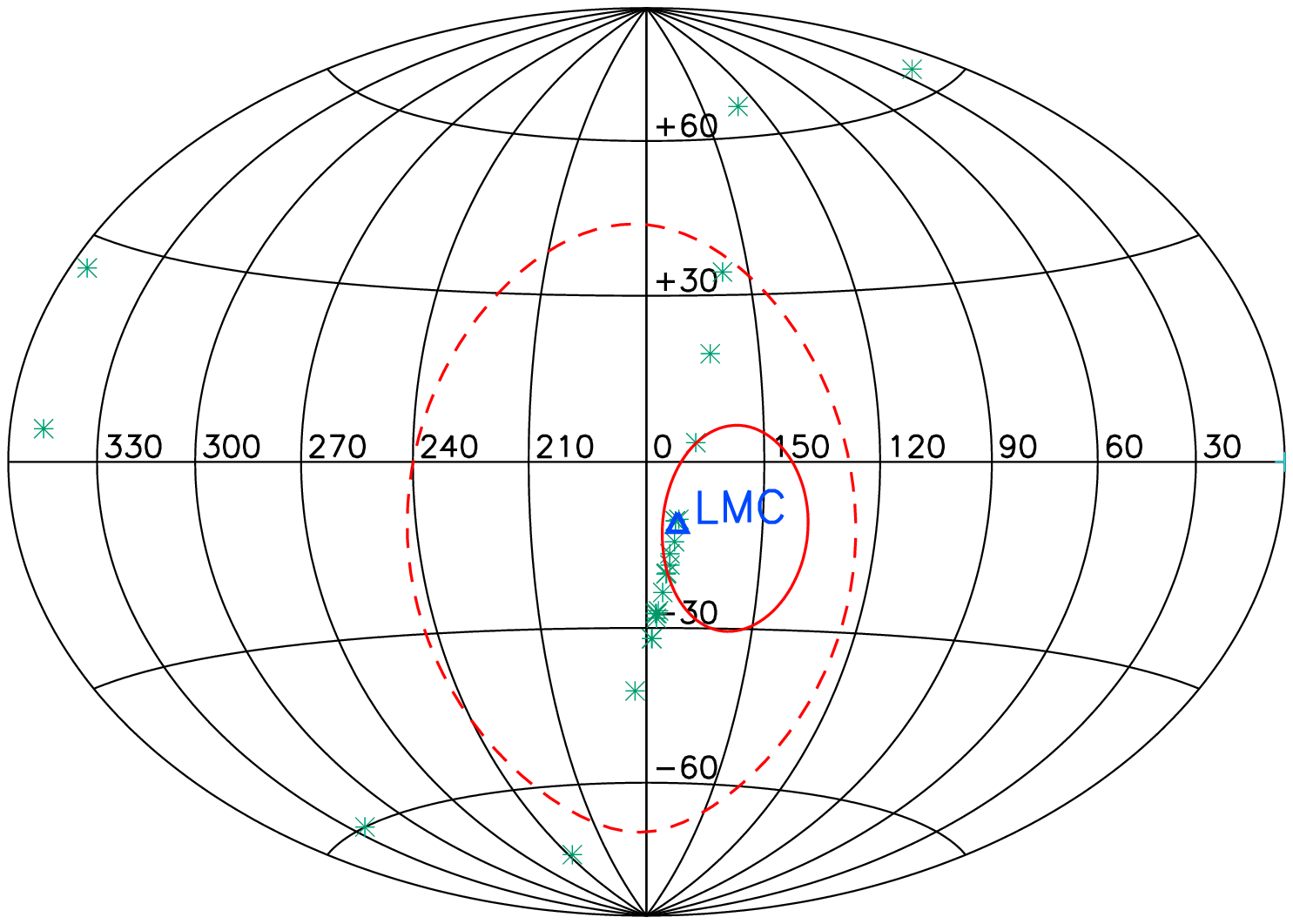} \\
    \end{tabular}
    \caption{Angular momentum directions of the selected particles for the small extension tidal tail (green asterisks) and the LMC (blue triangle). The dashed ellipse represents the uncertainty associated to the mean angular momentum of the 11 classical dwarf galaxies, while the solid ellipse represents the uncertainty associated to the direction perpendicular to the VPOS.}
    \label{fig:AM_DoS_simu}
  \end{center}
\end{figure*}

Let us now consider the first-passage tidal tail, which is the best candidate for inducing TDGs passing near the MW. Figure \ref{fig:TT_simu} shows the formation of clumps due to gravitational instabilities in the gaseous tidal tail at discrete locations. The five prominent over-densities have mass profiles consistent with that of dwarf galaxies with half-mass radii from 0.9 to 4 kpc, and total masses from 1.7 to 8.6 $\times$10$^8$ M$_{\odot}$ (see Table \ref{tab:TDGs}). Our 2.4 million particles simulation generates TDGs similar in location, e.g. the so-called "beads on a string" appearance, to those found in the \citet{Wetzstein2007} EG2 model with 6 million particles, but with larger masses. This is surprising, since the larger the resolution in the simulation, the larger the mass of the formed TDGs (see Sect. \ref{sec:model}). As suggested by \citet{Wetzstein2007}, this could be due to the initial gas fraction, which is twice in our simulation than the 30\% adopted in the \citet{Wetzstein2007} EG2 model and could result in the formation of more massive TDGs. However, the simulation does not have a resolution large enough to sample the lowest mass dwarf galaxies (see Table \ref{tab:TDGs}). A higher resolution and a smaller softening length is required to generate such TDGs \citep{Bournaud2008}, and one may expect a larger number of TDGs with masses similar to those of the MW companions.

Given our absence of knowledge about the detailed internal structure of the tidal tail, we have arbitrarily selected particles within a 40 kpc section of the tidal tail (300 particles) at the region interacting with the MW (see red arrow in Fig. \ref{fig:TT_simu}). In Fig. \ref{fig:DoS_simu} for the sake of the visualization, we have randomly selected only 25 particles amongst the 300 particles. They are all included into a thick plane, whose orientation is found to be similar to the VPOS within 13 degrees, a value smaller than the observational uncertainty (see Table \ref{tab:fitDoS1}). Finally, almost all their angular momentum directions are found to be close to the direction perpendicular to the observed VPOS, except for some of them which have an opposite direction as Sculptor (see Fig. \ref{fig:AM_DoS_simu}).

However even if this tidal tail give some promising results, it does not match the spatial distribution extension of the 11 classical dwarf galaxy (see Fig. \ref{fig:DoS_simu}, top panels). We notice that in the simulations reproducing M31 the first tidal tail may vary by factor 1 to 5 in width. This variation is due to the sensitivity to the initial conditions: the size of the gas disk in the progenitors and the orbital parameters. For example, orbital parameters associated to conditions close to the resonance (between the disk rotation and the orbital motion) at the first passage provide very narrow first tidal tails. Because being close to the resonance is not a requirement to the M31 model (in fact such conditions are mostly required at the second passage), we have used another model of M31 to generate a wider tidal tail. As the clump locations within the tidal tail have no other constrains than the actual spatial distribution of the classical dwarf galaxies, we have used three clumps instead of one to mimic the extension of the tidal tail and the results are shown in Figure \ref{fig:DoS_simu} (bottom panels).

We emphasize that this model is in sharp contrast with Sect. \ref{sec:mod1} and \citet{Pawlowski2011} studies that use all free parameters to optimize the tidal tail. In other words, the formation of the VPOS is predicted by the merger model of M31, as shown in Fig. \ref{fig:TT_IC}.

\begin{table}
  \caption{Physical parameters of the five TDGs identified in the tidal tail generated during the 1st passage between the M31 merger progenitors. Masses are given in unit of $10^{8}$ M$_{\odot}$. \emph{From left to right}: gaseous mass, stellar mass, total mass, gas fraction, and half mass radius (kpc).}
  \begin{center}
    \begin{tabular}{cccccc}
      \hline \hline
      No. &  $M_{\rm gas}$  &   $M_{\rm Star}$  & $M_{\rm Total}$  & $f_{\rm gas}$ & $r_h$ \\ %\tablenotemark{a} \\
          &  ($10^{8}$ $M_{\odot}$) & ($10^{8}$ $M_{\odot}$) & ($10^{8}$ $M_{\odot}$) &      & (kpc)$^{a}$  \\ \hline
      1   &  1.67 & 0.056   & 1.72   &   0.97       &  4.3  \\
      2   &  6.20 & 2.41    & 8.62   &   0.72       &  1.5  \\
      3   &  3.49 & 0.027   & 3.76   &   0.93       &  1.1  \\
      4   &  4.11 & 2.58    & 6.70   &   0.61       &  0.75 \\
      5   &  4.09 & 2.93    & 7.02   &   0.58       &  0.85 \\
      \hline
    \end{tabular}
  \end{center}
  \vspace{0.2cm} $^{a}${$r_h$ is the half mass radius of the TDG  for the 2D projected mass profile.}
  \label{tab:TDGs}
\end{table}

\begin{figure}
  \centering
  \includegraphics[width=0.8\linewidth]{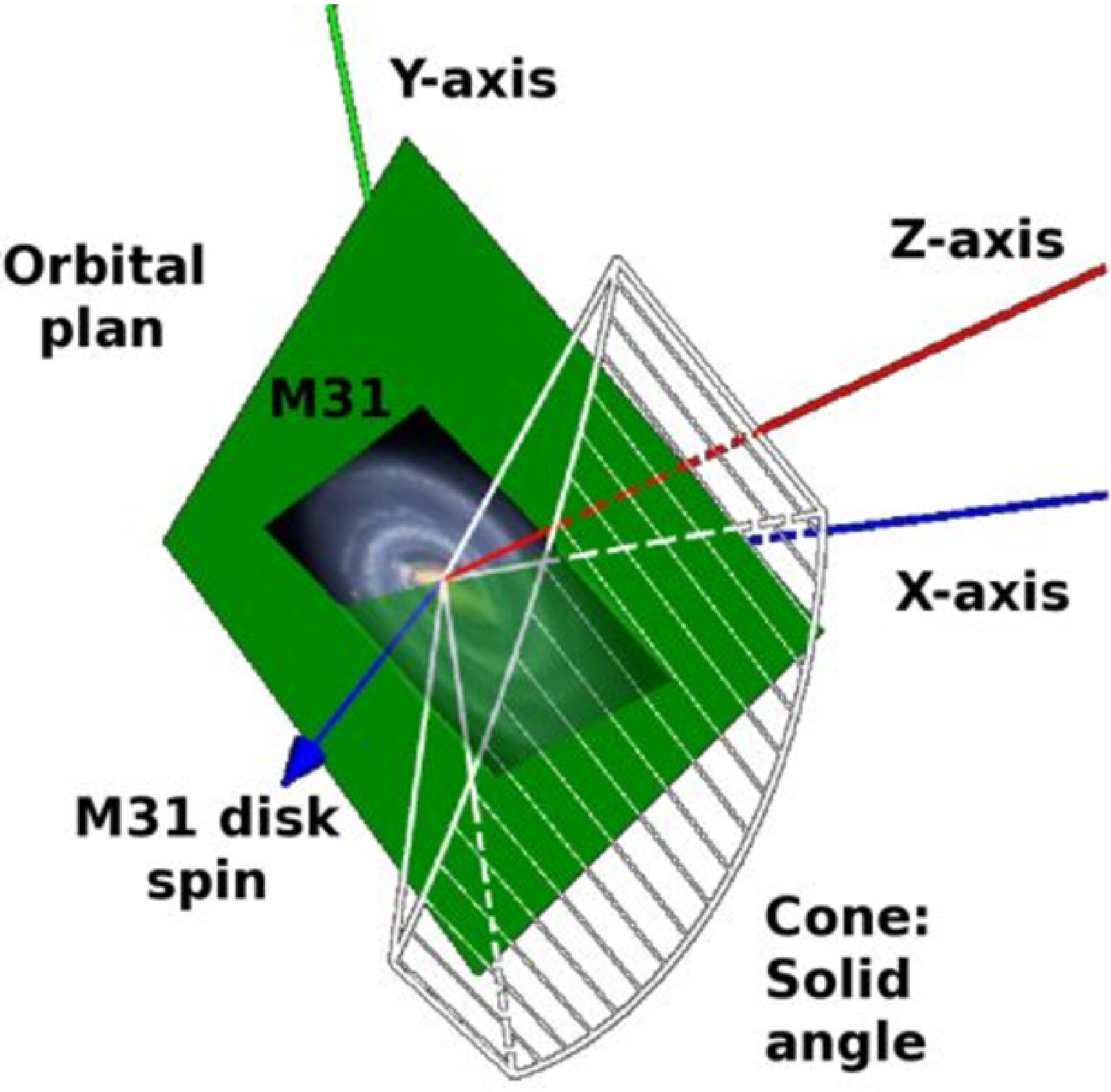}   
  \includegraphics[width=0.8\linewidth]{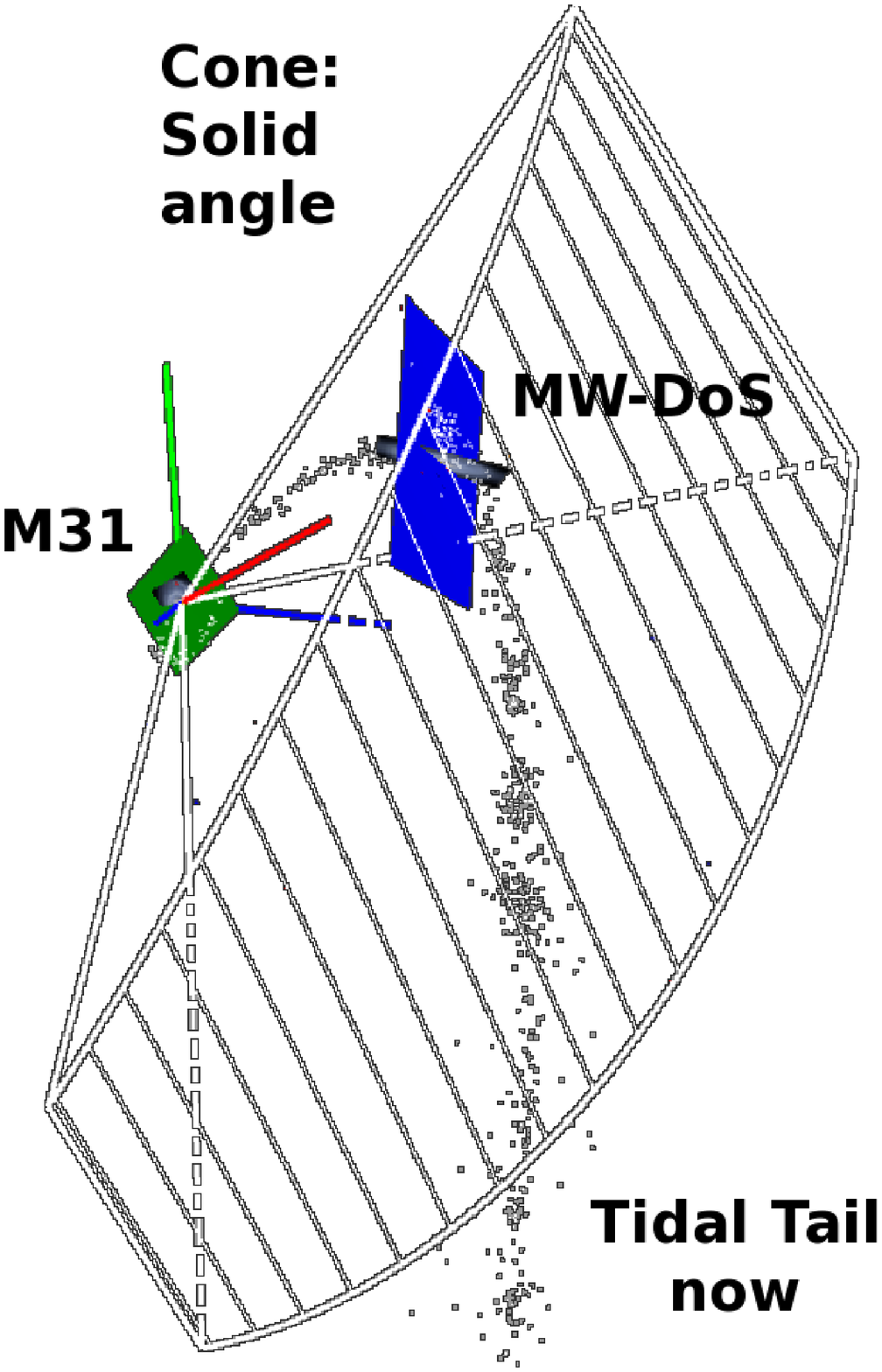}
  \caption{Geometrical view of the merger remnant. \emph{Top panel:} The disk plane of M31 is represented using a picture of a spiral galaxy with its spin direction shown as a blue arrow, while the orbital plane of the progenitors as shown in green. The white cone represents the solid angle that could be swept by the tidal tail ejected during the merger. \emph{Bottom panel:} Zoom on the solid angle swept by the tidal tail. The MW position and the plane of VPOS is shown in blue. The opening angle of the cone is constrained by the positions of TDG1 and TDG5 (see Fig. \ref{fig:TT_simu}) and the modeling of M31 and the Giant Stream. It accounts for uncertainties related to the modeling (varying parameters, see the text) as well as for a rotation around the M31 disk by $\pm$20 degrees along the M31 disk and $\pm$5 degrees in the plane of the tidal tail.}
  \label{fig:cone}
\end{figure} 

\section{Discussion}
\subsection{A surprising series of "coincidences" supporting an M31 origin}

At first sight, a possible causal link between a possible merger in the history of M31 and the MW dwarf galaxies might appear at odds with many years of studies of the Local Group. To our knowledge such a proposition has never been made, probably because it appears far too unlikely for geometrical reasons: why the MW would be precisely at a {\it rendez vous} with a tidal tail formed during the first passage, 8 to 9 Gyr ago, of a merger that occurred at more than 780 kpc away?

Let us consider three galaxies of similar masses, two of them being involved in a merger process. What is the chance that a tidal tail produced during the event reaches the third galaxy? Assuming a merger in the history of M31, it should be rather small given the fraction of the volume swept by a tidal tail. The corresponding conic volume is shown in Fig. \ref{fig:cone}, drawing a solid angle of 2000 square degrees, i.e, 5.5\% of the 4$\pi$ steradian sphere. In principle this argument should suffice to falsify or reject a possibility to associate M31 to the MW dwarf galaxies. In practice, the MW lies precisely within this small volume, because both the M31 disk is seen edge-on and the geometry of the Giant Stream (and of the Northern Loop) fixes the progenitors inclinations, and then the properties of the tidal tail induced after the first passage. We have considered a very large number of models from \citet{Hammer2010} as well as more recent ones. In these models, $r_{peri}$ ranges from 24 to 30 kpc, the mass ratio from 2.5 to 3.5, the dark to baryonic matter ratio from 10 to 25\%, and a significant range of initial progenitor inclinations are explored, which result in an uncertainty on the angular location of the first tidal tail that is smaller than 20 degrees. Of course we cannot claim having investigated all the numerous parameters of a merger model for M31, although the number of constraints just describing the M31 galaxy and its outskirts are quite impressive  \citep{Hammer2010}. We have also verified that for all geometrical configurations studied by \citet{Barnes2002}, the disk plane is close within less than 25 degrees to the orbital plane. This favorable configuration applies well to the M31 disk (see Fig. \ref{fig:cone}), which is supported by an angular momentum that is found to be slightly larger than the average of similar local spiral galaxies \citep[see][]{Hammer2007}.

Perhaps all of the above is just a coincidence and other properties would easily falsify a causal link between the M31 merger and the MW dwarf galaxies. Fig. \ref{fig:TT_IC} shows that the tidal tail velocity matches well that of the LMC, in amplitude and orientation. This is ensured using the most recent results compiled by \citet{Vieira2010} with quite large uncertainties (see their Fig. 10), while the amplitude of the LMC velocity would be too large using the \citet{Kalli2009} values. Such a match with the LMC velocity is another intriguing coincidence. While \citet{YY2010} have shown that the LMC past trajectory may have passed near M31 for a precise range of M31 transverse velocity, why would the velocity of the modeled tidal tail follow such an amplitude and orientation? This coincidence naturally ensures that our model can reproduce the VPOS orientation and dwarf angular momenta (see Fig. \ref{fig:DoS_simu}). 

Possibly the most astonishing, fully unexpected coincidence is the fact that the part of the tidal tail shown in Fig. \ref{fig:TT_simu} which interacts with the MW is very close to the most massive simulated tidal dwarf, TDG2, whose mass almost reaches that of the LMC. This is very encouraging for pursuing this study with higher resolution simulations.

\subsection{Could the MW dwarf galaxies be descendants of TDGs?}
\subsubsection{Global properties}

First we acknowledge that a robust proof of a dominant dark matter component in MW dwarf galaxies would provide a definitive falsification of the scenario proposed in this paper. It may come from the detection of dark matter annihilation in MW dwarf galaxies. In fact the non-detection of dark matter annihilation in MW dwarf galaxies could provide the strongest up to date constraint on the nature of the dark matter particles, excluding for some annihilation channels, thermal dark matter for masses below 30 GeV or so  \citep[see e.g.,][]{Ackermann2012}. On the other hand, a tidal origin for the MW dwarf galaxies would discard a significant dark matter content in these galaxies. It was considered very early \citep[e.g.,][]{Lynden-Bell1976} and was revived by \citet{Kroupa1997,Kroupa2005}. Supports for this assumption are numerous and well summarised by \citet{Kroupa2010} and \citet{Kroupa2012}. Their old stellar population are perhaps inherited from the merger progenitors. Their long lifetime in the tidal tail could have let them in various environments providing a diversity in star formation histories that could explain their different behaviours. \citet{Kroupa2010} summarised that "the physics of TDG formation and evolution is sufficiently well understood to conclude that 1) once formed at a sufficient distance from the host, TDGs will take an extremely long time to dissolve, if at all; and 2) the TDGs formed will naturally lead to a population of ancient TDGs that resemble dSph satellites".

However to be plausible our scenario \citep[as well as that of][]{Pawlowski2011} requires an alternative explanation of the large mass-to-light ratios of the MW dwarf galaxies. Recently \citet{Casas2012} have simulated dark-matter free satellites that after several orbits around the MW are out-of-equilibrium bodies with high apparent mass-to-light ratios. According to their claim, if such progenitors of MW satellites have reached the MW potential at  the same time along a few specific orbits, they may reproduce most of their intrinsic kinematics and surface brightness. Although similar simulations are beyond the scope of this paper, we noticed that the \citet{Casas2012} progenitors are far more compact than the TDGs that are formed from our modelling. In other words, it is likely that realistic TDGs are more fragile and they may be even more easily disrupted than in the \citet{Casas2012} study. In a forthcoming paper we will study a more realistic TDG infalling into the MW potential and to verify whether or not such an event may reproduce the MW dwarf galaxies. Such a study is mandatory to verify whether or not the large mass-to-light ratios of the MW dwarf galaxies falsify the M31 scenario.

\subsubsection{Could the LMC be descendant of a TDG?}

The above mentioned studies do not apply to the Magellanic Clouds and one may wonder whether or not the most massive Irr galaxies of the Local Group could also be TDGs. In the M31 scenario, both LMC and SMC have arrived for the first time in the MW halo, and as such they are not stripped of their gas, conversely to other MW dwarf galaxies. The stellar mass of the LMC has been properly estimated by \citet{VdM2002} assuming an LMC V-band absolute magnitude of $M_{\rm V}=-18.5$, an extinction of $A_{V}$=0.4, and an extinction-corrected color $(B-V)_0 = 0.43$. Following \citet{Bell03}, the stellar mass for the LMC is estimated to be $2.6 \times 10^{9} M_{\odot}$ for a diet Salpeter IMF. Based on a full reconstruction of the star formation history, \citet{Harris2009} found a stellar mass ranging from 1.7 to 3.5$ \times 10^{9} M_{\odot}$ (see their Fig. 12) for a Salpeter IMF. To be compared with a diet Salpeter IMF value, one has to subtract 0.15 dex \citep{Bell03} to these values, reaching a range of 1.2 to 2.5 $\times 10^{9} M_{\odot}$. Other IMFs generally provides a lower stellar mass, by subtracting an additional 0.15 dex for e.g. a \citet{Kroupa2002} or a \citet{Kennicutt1983} IMF. According to \citet{Kim1998}, the LMC gas mass is $0.5  \times10^{9} M_{\odot}$ and thus the baryonic mass ranges from 1.7 to 3.1 $\times 10^{9} M_{\odot}$, assuming a diet Salpeter IMF for the stellar mass estimate. These values are larger than those of Table \ref{tab:TDGs} by a factor 2 to 3, which could be a serious difficulty for a tidal origin of the LMC. More detailed and resolved simulations of gas-rich tidal tail with more realistic physical conditions may help to further investigate this issue. On the observational side, \citet{Kaviraj2011} found several TDGs (approximately 15\% of them) with a mass larger than 10$^9$ M$_{\odot}$, after a wide TDG search in the SDSS data through the Galaxy Zoo project. The fact that these local mergers are certainly less gas-rich than the assumed ancient M31 merger progenitors suggests the latter may lead to TDGs as massive as the LMC.

To investigate whether a tidal origin related to a merger at the M31 location is realistic or not would require to compare it with the numerous results provided by studies on the Local Group and dwarf galaxies.  It also requires to firmly assess the model properties and its predictions before any conclusion. For example the recent determination of the Fe/H abundance of old stars in the LMC may be not a falsification of this scenario \citep[see e.g.][]{Haschke2012}. The fact that the old globular clusters in M31 are relatively Fe/H richer can be explained by our model. The LMC is predicted to be formed within the first passage tidal tail with stars stripped from the less massive interloper 8 to 9 Gyr ago, while old globular clusters could have been formed from material stripped from the most massive interloper, i.e., expected to be metal richer. The search for further falsification tests has to continue.

\subsubsection{MW dwarf velocities and angular momenta}

\begin{figure}
  \centering
  \includegraphics[width=1\linewidth]{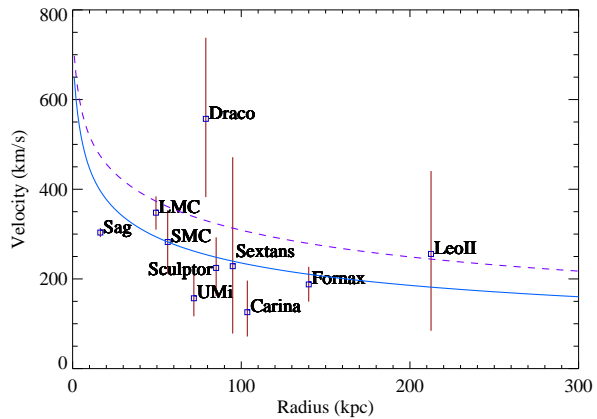}
  \caption{MW escape velocity as a function of radius for two baryonic-to-total mass ratios: 6\% (magenta dashed line) and 20\% (blue line). The baryonic mass remains equal to 6.6$\times$10$^{10}$ M$_\odot$ in both cases. The red lines represent the uncertainty associated with the velocities of the classical dwarf galaxies.}
  \label{fig:fig_VR}
\end{figure}

It is unclear yet whether all classical dwarf galaxies are part of the VPOS. Sagittarius is clearly off the VPOS in Fig. \ref{fig:AM_obs}. Could it be due to the fact that it is strongly tidally disrupted by the MW potential? Figure \ref{fig:fig_VR} provides a crucial test of the M31 scenario by comparing the dwarf velocities to the escape velocities assuming two values for the MW mass. However this exercise is limited by the accuracy of the proper motion data: this excludes Draco, Sextans and LeoII for which the uncertainties are so large that their bound or unbound nature remains unknown. For example, the Sextans proper motion was not measured but derived, and the Draco proper motion was calculated by a ground-based telescope more than 10 years ago providing an extremely high value, $V$ = 556 km/s. It results that Sag, UMi and Carina are certainly bound for any values of the MW mass, while all other dwarf galaxies are lying in a region where the measurement accuracy does not allow us to conclude. Perhaps the Sag, UMi and Carina properties are not consistent with the M31 merger scenario, and only much more accurate measurements of proper motion will allow to provide a similar test for the other dwarf galaxies. Using proper motions to distinguish between the various scenarios for the formation of the dwarf galaxies is very tempting but often limited by the data accuracy \citep[see e.g.,][]{Angus2011}, and by the poor knowledge about the true MW total mass. Moreover this also depends on the nature of their progenitors. If assumed to be of tidal origin from a gas-rich merger, they should be gas-rich as shown in Table \ref{tab:TDGs} prior they interact with the MW potential. 

Possibly, when entering the MW potential, these gas-rich TDGs are entirely stripped of their gas through tidal shocks and ram pressure stripping as described by \citet{Mayer2007}. More recently, many authors \citep[see e.g.,][and references therein]{Mayer2010,Mayer2011} have shown that tidal stirring is an efficient mechanism to remove the gas from gas-rich, dark matter dominated, dwarf irregular galaxies interacting with the potential of a massive galaxy such as the MW.  This leads to their transformation into dwarf spheroidals, which turns rotationally supported stellar systems into pressure supported ones. The interplay between dynamical friction and tidal mass evaporation not only decreases strongly their mass by factors reaching 10, it affects also their angular momentum and velocity as shown by \citet{Taffoni2003}. This might lead to a $\sim$ 20\% decay in velocity during 2 Gyr (see their Fig. 6), e.g., providing initial velocities of UMi and Carina consistent with those of other dwarf galaxies. Our initial guess is that tidal stirring should be more efficient for progenitors that are not dark matter dominated such as TDGs. These are however pure conjectures that underlines the severe limitations of our modeling for which we have represented tidal tail particles as point masses. For reaching a conclusive answer about the consistency of the M31 merger scenario with the dwarf properties would require a far more complete modeling, perhaps taking into account the interaction between the whole tidal tail and its associated TDGs with the MW potential. Up to our knowledge, such a simulation have never been done by assuming a dark-matter free irregular galaxy such as a realistic TDG falling in the potential of a massive galaxy. We are encouraged to perform this in a future study also because we notice that the most inconsistent galaxy velocities (those of Sag, UMi and Carina) have also the angular momenta the most shifted from the averaged dwarf galaxy angular momentum (see Fig. \ref{fig:AM_obs}).
 
\section{Conclusion: a causal link between the M31 major merger and the VPOS?}

It is widely accepted that the classical bulge of M31 is due to an ancient major merger \citep{VdB2005, KK2004}. Quoting John Kormendy (2010, private communication): "don't we already know with some confidence that M31 was formed in at least one (in fact: very probably exactly one) major merger?" Perhaps its consequences have not been fully investigated, whereas it occurred in the galaxy that represents almost two third of the Local Group baryonic mass. The immediate question is: where are the relics of the most energetic event that occurred in the whole Local Group history? One may investigate the most anomalous and difficult to explain features discovered in the Local Group. Among them are the M31 haunted halo \citep{Ibata2007}, the Giant Stream \citep{Ibata2001}, the VPOS and also the unusual proximity of two massive Irr  near the MW \citep[see e.g.,][]{VdB2010}. 

In principle a $\sim$ 1 Gyr old minor merger may reproduce the Giant Stream \citep[e.g.][]{Fardal2008} although this may be problematic \citep[see e.g.][]{Font2008} because stars in the Giant Stream have ages from 5.5 to 13 Gyr \citep{Brown2007}.  \citet{Hammer2010} have been able to reproduce the Giant Stream as well, assuming a major merger and using the stellar ages in the M31 outskirts as clocks to date a major merger, providing a first passage and a fusion occurring $8.75 \pm 0.35$ Gyr and $5.5 \pm 0.5$ Gyr ago, respectively. Having M31 involved in a merger at these epochs may be quite common, as demonstrated by detailed studies of intermediate redshift galaxies from the IMAGES sample \citep{Hammer2009,Puech2012}. This model is quite challenging in reproducing many other properties of the M31 galaxy (disk, bulge, thick disk and 10 kpc ring) together with the Giant Stream. A noticeable feature of gas-rich major mergers is that they produce long-lived tidal tails that can reproduce many structures discovered in the halo of giant spiral galaxies \citep[see e.g.,][]{Wang2012}.

This paper is the first of a series aiming at investigate consequences of such a major merger occurring in the Local Group. The main weakness of the present modeling is due to the oversimplification in considering tidal tail particles as point masses. To reproduce the properties of the MW dwarf galaxies requires a full modeling of extended, gas-rich TDGs entering into the potential of the MW. It is highly desirable to verify whether or not the MW dwarf galaxies and the VPOS can be fully modeled by the interaction between a gas-rich tidal tail, its associated TDGs, and the MW. Such a study will aim at verifying whether or not the intrinsic properties of dwarf galaxies may be reproduced, including their apparent large mass-to-light ratios.

Our modeling of the M31 major merger is getting more and more mature since \citet{Hammer2010} and it is able to provide some predictions that can be tested for falsification purposes. The modeling of an anisotropic feature like the Giant Stream, and further on, of the Northern loop (Yang et al. 2012, in prep.) almost fix the location and velocity of the tidal tails associated to the merger. Here we have concentrated our efforts to examine the largest tidal tail, which is formed during the first passage, stripping the minor encounter stars, 8 to 9 Gyr ago. This tidal tail is bringing the largest amount of baryonic material expelled by this event towards the Local Group. It results in a series of "coincidences" that considerably strengthens our preliminary conjecture. The MW is found precisely at the meeting point with the tidal tail, at present time. Moreover, the velocity of the tidal tail matches well that of the LMC providing a good fit of the VPOS, including its spatial orientation and the angular momentum, even to some details such as the inverted angular momentum of Sculptor. 

Having this in mind we conclude that a link between the VPOS and the major merger at the M31 location is plausible. Compared to the MW merger or encounter hypothesis of \citet{Pawlowski2011}, it has the advantage of resulting from a testable prediction that could be falsified or supported by many observations of the Local Group. Besides this, the present modeling is made within the $\Lambda$CDM scenario and there is no peculiar need to account for other physics to describe the above noticed unusual properties of the Local Group, i.e. the VPOS, the presence of two massive Irr near the MW and most features found in the M31 halo. In fact, \citet{Knebe2011} have been the first to show that within $\Lambda$CDM, there should be renegade sub-halo, i.e. sub-halo of M31 that could have migrated towards the MW. However they could not explain neither the VPOS nor the LMC and its large velocity, conversely to our prediction that lead to an external origin for most MW dwarf galaxies. If the latter is true, it strengthens the tension between the predicted number of dark matter sub-haloes and observations, i.e. the so-called missing satellite problem, which might be then reconsidered as an excess of sub-haloes in the $\Lambda$CDM paradigm.

Note added into proofs : \cite{Pawlowski2012b} recently shows that filamentary infall is excluded to explain the VPOS.

\section*{Acknowledgments}
This research has made use of the NASA/IPAC Extragalactic Database (NED) which is operated by the Jet Propulsion Laboratory, California Institute of Technology, under contract with the National Aeronautics and Space Administration. We are grateful to Jianling Wang for generating a part of the models of M31 that have been used in this study. Finally, to represent the Local Group in 3D, three-dimensional visualization was conducted with the S2PLOT programming library. We warmly thanks the referee whose remarks have significantly improved the manuscript including by identifying the important tests that should be made to put the M31 scenario in a cosmological context.

\appendix
\section{Statistical method to test the isotropy}

The statistical test proposed by \citet{Metz2007a} assesses whether the MW dwarf galaxy positions or velocities derive from an isotropic distribution. 50 000 random samples are generated using Monte-Carlo simulations. Each sample is isotropic and follows a radial density distribution $\rho(r) \propto 1/r^2$, consistent with the radial distribution of the MW dwarf galaxy positions \citep{Kroupa2005} and velocities. A plane is fitted for each sample, and the associated uncertainty is estimated using bootstrap resampling as follows. Each sample is resampled 1~000 times and the perpendicular vector for each of the corresponding fitted plane is stored in a matrix M. The 3D symmetric matrix $S = M^TM$ is computed and its three eigen values ($\tau_1 < \tau_2 < \tau_3$) calculated. To characterize the 1 000 vector distribution, i.e., to characterize the plane fit uncertainty, two numbers are defined using the eigen values:
$$\gamma = \frac{\log{(\tau_3 / \tau_2})}{\log{(\tau_2 / \tau_1})}$$
$$\zeta  = \log{(\tau_3 / \tau_1)}$$
$\gamma$ describes how clustered the vector distribution is, while $\zeta$ indicates how the vector distribution focuses towards one direction. If $\gamma$ and $\zeta$ are small ($\sim 1$) the vector distribution is close to isotropic (the plane fit uncertainty is thus large), whereas if $\gamma$ and $\zeta$ are large ($> 2$) the vector distribution clearly defines one direction (the plane fit is then well defined). The resulting statistics of $\gamma$ and $\zeta$ is plotted in Fig. \ref{fig:AM_obs}. For the 10 classical dwarf galaxy position and velocity set, $\gamma$ and $\zeta$ are also derived. The distribution function D($\gamma_0$, $\zeta_0$), defined by the probability that $\gamma > \gamma_0$ and $\zeta > \zeta_0$, can be used to estimate the probability for a sample to derive from an isotropic law.

\bsp

\label{lastpage}

\end{document}